\documentclass[journal,onecolumn]{IEEEtran}
\IEEEoverridecommandlockouts
\usepackage{cite}
\usepackage{amsmath,amssymb,amsfonts}
\usepackage{graphicx}
\usepackage{subcaption}
\usepackage{color, xcolor}
\usepackage{multirow}
\usepackage{nicematrix}

\usepackage{amsthm}
\usepackage{bm}
\usepackage{blkarray}
\usepackage{algorithm}
\usepackage{enumitem}
\newtheorem{definition}{Definition}
\newtheorem{example}{Example}
\newtheorem{construction}{Construction}
\newtheorem{corollary}{Corollary}
\newtheorem{theorem}{Theorem}
\newtheorem{lemma}{Lemma} 
\newtheorem{remark}{Remark}

\begin{document}
	\title{A New Construction Structure on MISO Coded Caching with Linear Subpacketization: Half-Sum Disjoint Packing}
	\author{
		Bowen Zheng, Minquan Cheng,~\IEEEmembership{Member,~IEEE, } 
		Kai Wan,~\IEEEmembership{Member,~IEEE,}  and Giuseppe Caire,~\IEEEmembership{Fellow,~IEEE}
		\thanks{B. Zheng and M. Cheng are with Guangxi Key Lab of Multi-source Information Mining $\&$ Security, Guangxi Normal University,
			Guilin 541004, China (e-mail:  zhengbowen9812@163.com, chengqinshi@hotmail.com).}
		\thanks{K.~Wan is with the School of Electronic Information and Communications,
			Huazhong University of Science and Technology, 430074  Wuhan, China,  (e-mail: kai\_wan@hust.edu.cn).}
		\thanks{G. Caire is with the Electrical Engineering and Computer Science Department, Technische Universit\"{a}t Berlin,10587 Berlin, Germany (e-mail: caire@tu-berlin.de).}
	}
	\date{}
	\maketitle
	
	\begin{abstract}
		In the $(L,K,M,N)$ cache-aided multiple-input single-output (MISO) broadcast channel (BC) system, the server is equipped with $L$ antennas and communicates with $K$ single-antenna users through a wireless broadcast channel where the server has a library containing $N$ files, and each user is equipped with a cache of size $M$ files. Under the constraints of uncoded placement and one-shot linear delivery strategies, many schemes achieve the maximum sum Degree-of-Freedom (sum-DoF). However, for general parameters $L$, $M$, and $N$, their subpacketizations increase exponentially with the number of users. We aim to design a MISO coded caching scheme that achieves a large sum-DoF with low subpacketization $F$. An interesting combinatorial structure, called the multiple-antenna placement delivery array (MAPDA), can be used to generate MISO coded caching schemes under these two strategies; moreover, all existing schemes with these strategies can be represented by the corresponding MAPDAs. In this paper, we study the case with $F=K$ (i.e., $F$ grows linearly with $K$) by investigating MAPDAs. Specifically, based on the framework of Latin squares, we transform the design of MAPDA with $F=K$ into the construction of a combinatorial structure called the $L$-half-sum disjoint packing (HSDP). It is worth noting that a $1$-HSDP is exactly the concept of NHSDP, which is used to generate the shared-link coded caching scheme with $F=K$. By constructing $L$-HSDPs, we obtain a class of new schemes with $F=K$. Finally, theoretical and numerical analyses show that our $L$-HSDP schemes significantly reduce subpacketization compared to existing schemes with exponential subpacketization, while only slightly sacrificing sum-DoF, and achieve both a higher sum-DoF and lower subpacketization than the existing schemes with linear subpacketization. 
	\end{abstract}
	
	
	\section{INTRODUCTION}
	With the exponential growth of data traffic and the proliferation of mobile devices, modern communication networks face increasing pressure to cope with the rising demand for low-latency and high-throughput content delivery. In particular, the high temporal variability of network traffic leads to congestion during peak periods and underutilization during off-peak periods. Coded caching, originally proposed by Maddah-Ali and Niesen (MN) in \cite{MN}, has emerged as a promising technique to address these challenges by effectively leveraging storage at the network edge and multicasting opportunities to improve transmission efficiency. 
	
	The $(K, M, N)$ MN coded caching problem was initially designed for the single-input single-output (SISO) shared-link network model, where a central server with $N$ files connects to $K$ cache-aided users through an error-free shared link\footnote{In this paper, we consider the case $N \geq K$.}. Each user can store up to $M$ files, where $0 \leq M \leq N$. An $F$-division $(K, M, N)$ coded caching scheme consists of two phases: the placement phase and the delivery phase. In the placement phase, the server divides each file into $F$ equal-size packets and places some packets into each user's cache without knowing their future requests. If the packets are directly placed in each user's cache, it is called uncoded placement. Otherwise it is called coded placement. The parameter $F$ is referred to as the subpacketization. In the delivery phase, each user independently requests a file. Based on these demands and the cached packets for each user, the server sends some coded packets such that each user can decode its requested file with help of its cached packets. 
	
	The transmission load (or load) $R$ is the amount of transmission for the worst demand case of the users, normalized by the size of a single file. The coded caching gain of a coded caching scheme is defined as $g=K(1 - M/N)/R$ which represents the average number of users served by the server at each time slot, where $K(1 - M/N)$ denotes the transmission load of a conventional uncoded caching scheme. The objective to design a scheme with the load $R$ as small as possible, i.e., the coded caching gain $g$ is as large as possible. 
	
	Maddah-Ali and Niesen in~\cite{MN} proposed the first well-known coded caching scheme, which is referred to as the MN scheme. When $t = KM/N$ is an integer, the MN scheme achieves the load  $R=(K-t)/(t+1)$, i.e., the coded caching gain $g=t + 1$. The authors in \cite{WTP} showed that the load is optimal under uncoded placement when $K \le N$. For arbitrary user demands, the authors in \cite{YMA} derived the converse of the transmission load under uncoded placement and proposed an optimal scheme by eliminating the redundant multicast transmissions in the delivery phase of the MN scheme. So, the MN scheme has been widely used in the variety of network topologies, such as Device-to-Device (D2D) networks~\cite{JCM}, hierarchical networks~\cite{KNMD}, combination networks~\cite{JWTLCEL}, multi-server linear networks~\cite{SMK}, etc.
	
	However, the MN scheme has the subpacketization $F = \binom{K}{t}$ which increases exponentially with the number of users $K$. There are many studies focusing on reducing the subpacketization by increasing the load, such as the placement delivery array (PDA)\cite{YCTC,CJYT,CJWY,CWZW,WCWC}, projective geometry\cite{CKSM}, hypergraphs~\cite{SZG}, Ruzsa-Szem\'{e}redi graphs\cite{STD}, strong edge coloring of bipartite graphs~\cite{YTCC}, other combinatorial designs~\cite{ASK}.
	
	The authors in \cite{SZG} pointed out that all the schemes in \cite{SZG, STD, YTCC, ASK} can be represented by appropriate PDAs, indicating that PDAs are a powerful tool for characterizing coded caching schemes. However, a PDA is merely a combinatorial structure that characterizes the placement and delivery issue in shared-link network, and its definition does not provide explicit guidance for constructing coded caching schemes. To establish a framework for constructing PDAs with linear subpacketization, the authors in \cite{CWWC} introduced a novel combinatorial structure, termed non-half-sum disjoint packing (NHSDP) which unifies placement and delivery strategies into a single condition. By constructing NHSDPs, a class of coded caching schemes with $F=K$ was obtained, which achieves a lower or only slightly higher load compared to some existing schemes under some system parameters.
	
	\subsection{MISO Coded Caching} 
	The $(K,M,N)$ MN coded caching problem was extended to the $(L,K,M,N)$ cache-aided multiple-input single-output (MISO) broadcast channel (BC) in  \cite{NMA} where the server equipped with $L$ antennas connects with $K$ single-antenna users over a wireless broadcast channel. The objective of the MISO BC problem is to maximize the sum-DoF (i.e., the average number of users simultaneously served per transmission block) of the system. When $L=1$, the sum-DoF is exactly the coded caching gain in the MN coded caching problem. Under the constraints of uncoded placement and one-shot\footnote{Each required data bit is transmitted exactly once in coded or uncoded form.} zero-forcing (ZF) linear delivery strategies, various schemes achieving the sum-DoF $\min\{L+t, K\}$ were proposed in~\cite{NMA, SCK, LE, SPSET, MB, ST, STSK}, and this sum-DoF was subsequently proved to be optimal under the above constraints in~\cite{LBE}. Specifically, the schemes in \cite{NMA, SCK} require a subpacketization on the order of $\binom{K}{t}\binom{K-t-1}{L-1}$. Subsequently, various studies have focused on reducing subpacketization while maintaining maximum sum-DoF in \cite{LE, SPSET, MB, ST, STSK} under the different system parameter constraints. For instance, when $K/L$ and $t/L$ are integers, the author in~\cite{LE} reduced the subpacketization to $\binom{K/L}{t/L}$; when $L \geq t$, the authors in~\cite{SPSET} further reduce the subpacketization to linear subpacketization under the cyclic placement. 
	
	For the general system parameters, the authors in \cite{YWCC, NPR} independently introduced a combinatorial structure, called the \textit{Multiple-antenna Placement Delivery Array} (MAPDA), which can be regarded as the generalization of the PDA, to study the low subpacketization. They also showed that all existing schemes in~\cite{NMA, SCK, LE, SPSET, MB, ST, STSK} can be represented by MAPDAs. Based on MAPDA, ~\cite{YWCC, NPR} proposed some new schemes with the optimal sum-DoF which have much smaller subpacketization than those of \cite{NMA} and \cite{SCK}. It is worth noting that the scheme in \cite{YWCC} generalizes both \cite{LE} (when $m=L$) and \cite{MB} (when $m=1$) and some other linear subpacketization schemes are obtained for the special system parameters in \cite{YWCC, NPR}. Based on MAPDA, the authors in \cite{WCC} obtained the scheme with sum-DoF $2L$ under cyclic placement. The schemes in \cite{CTWWL} further improves the sum-DoF in \cite{WCC} and  generalizes \cite{SPSET}. A detailed summary of these existing schemes and their parameter limitations is listed in Table~\ref{table_existing_schemes}.
	
	In Table~\ref{table_existing_schemes}, we can see that the schemes in \cite{YWCC, NPR} achieve the optimal sum-DoFs while the subpacketizations grow exponentially with the number of users, or can achieve the optimal performance only in the special case where $t+L \geq K$. The schemes in \cite{WCC,CTWWL} achieve linear subpacketization but the sum-DoFs are just $2L$ when $t>L$ and $K>2t$. When $K$ is large and the memory ratio is small, the case $t>L$ and $K>2t$ always holds for any give parameter $L$. 
	
	Similar to the role of PDA in shared-link networks, MAPDA serves as a combinatorial structure that characterizes the placement and delivery strategies in the MISO coded caching system, but its definition alone does not provide explicit guidance for constructing coded caching schemes with both low subpacketization and large sum-DoF. So it is meaningful to propose frameworks for constructing MAPDAs with linear subpacketization.
	
	\begin{table}[htbp!]
		\centering
		\caption{The existing schemes with $L$ antennas and memory ratio of $M/N = t/K$ where $t \in [K]$, $\beta = \gcd(K, t, L)$, $\langle K \rangle_t = K \bmod t$, $t + L \leq K$, }
		\label{table_existing_schemes}
		\renewcommand{\arraystretch}{2.5}
		\begin{tabular}{|c|c|c|c|}
			\hline
			Scheme & Subpacketization & Sum-DoF $g$ & Limitation \\ 
			\hline
			\multirow{2}{*}{YWCC $1$ \cite{YWCC}} & $\frac{t+L}{\gcd(m, L-m)} \binom{K/m}{t/m}$ & \multirow{2}{*}{$t+L$} & $K/m$, $t/m \in \mathbb{Z}^+$, $m<L$  \\ \cline{2-2} \cline{4-4}
			& $\binom{K/L}{t/L}$ & & $K/L$, $t/L \in \mathbb{Z}^+$ \\ \hline
			YWCC $2$ \cite{YWCC} & $K$ & $t+L$ & $t+L \geq K$\\ \hline
			NPR \cite{NPR} & $\frac{t+L}{\beta}\binom{K/\beta}{(t+L)/\beta}$ & $t+L$ & $K/\beta$, $(t+L)/\beta \in \mathbb{Z}^+$\\ \hline
			\multirow{3}{*}{WCC\cite{WCC}} & $2LK$ & \multirow{3}{*}{$2L$} & $2 \nmid (K-t)$ \\ \cline{2-2} \cline{4-4}
			& $LK$ &  & $2 \mid (K-t)$, $L \nmid K$ \\ \cline{2-2} \cline{4-4}
			& $K$ & & $2 \mid (K-t)$, $L \mid K$ \\   \hline
			\multirow{2}{*}{CTWWL\cite{CTWWL}} & \multirow{2}{*}{$\frac{gK}{\beta^2}$} & $2L\lfloor \frac{t+L}{K-t+L} \rfloor + \langle t+L \rangle_{K-t+L}$ & $L \leq \langle t+L \rangle_{K-t+L} < 2L$ \\ \cline{3-4}
			&  & $ 2L\lfloor \frac{t+L}{K-t+L} \rfloor + 2L$ & $2L \leq \langle t+L \rangle_{K-t+L}$ \\  \hline
		\end{tabular}
	\end{table}
	
	\subsection{Research Motivation and Contribution}
	In this paper, we focus on designing MISO coded caching schemes with linear subpacketization, aiming to maximize the sum-DoF under uncoded placement and zero-forcing one-shot linear delivery strategies when $N \geq K$. 
	Through the analysis of Table~\ref{table_existing_schemes}, it can be seen that for existing MISO coded caching schemes with linear subpacketization, optimal sum-DoF is achievable only when $KM/N\leq L$, whereas in the regime $L<KM/N$, there remains significant room for improvement in both sum-DoF and subpacketization.
	To this end, we propose a novel combinatorial structure termed $L$-half-sum disjoint packing (HSDP), which generalizes the NHSDP introduced in \cite{CWWC} for shared-link systems to the MISO BC setting. 
	Note that the $1$-HSDP is exactly the NHSDP.
	Compared with existing characterizing methods, the main advantage of $L$-HSDP is that it unifies placement and delivery strategies into a single condition, i.e., the $L$-half-sum condition in Definition~\ref{def-LHSDP}.
	More precisely, the main contributions are summarized as follows:
	\begin{itemize}
		\item We introduce a unified framework for designing MISO coded caching schemes with linear subpacketization based on the $L$-HSDP. 
		By exploiting the structure of cyclic Latin squares, an $L$-$(v,g,b)$ HSDP for any odd positive integers $v$, $g$, and $b$ leads to an $(L,K=v,M,N)$ MISO coded caching scheme with memory ratio $M/N = 1 - bg/v$, subpacketization $F = K$, and sum-DoF $g$. 
		We can obtain a scheme with even number of users $K=v-1$ by considering one user as the virtual user in the aforementioned scheme.
		
		\item 
		We propose a construction framework for $L$-HSDPs based on embedding integers into a high-dimensional geometric space. 
		Unlike the construction of NHSDPs, the core idea in constructing an $L$-HSDP is to design a subset $\mathcal{X} \subset \mathbb{Z}_v$ so as to obtain an $L$-HSDP $(\mathbb{Z}_v,\mathfrak{D})$ in which every integer in each block of $\mathfrak{D}$ can be represented by $\mathcal{X}$, while the frequency of such representations is constrained by Condition~2 of Definition~\ref{def-LHSDP}. 
		In contrast, in an NHSDP, each integer in every block admits a unique representation via $\mathcal{X}$.
		It is worth noting that in a high-dimensional geometric space, designing a subset mapped to $\mathcal{X}$ for an NHSDP is equivalent to finding a collection of special linearly independent vectors, whereas designing a subset mapped to $\mathcal{X}$ for an $L$-HSDP corresponds to finding a collection of special linearly dependent vectors. Moreover, our objective is to maximize the number of such vectors, since this minimizes the memory ratio of the generated scheme for a given sum-DoF.
		Clearly, for a given odd integer $v$ and sum-DoF, constructing an $L$-HSDP is much more challenging than constructing an NHSDP.
		\begin{itemize}
			\item For any positive integers $m_1, m_2, \ldots, m_n$ with $L \leq 2^r$, by recursive construction and solving the linear subpaces we obtain our desired $\mathcal{X}$ and construct a $L$-HSDP which generates an $(L,K=v,M,N)$ MISO coded caching scheme with memory ratio $M/N = 1-(2^{n+r}\prod_{i=1}^{n}m_i)/v$, subpacketization $F = K$, and sum-DoF $g = 2^{n+r}$, for any odd integer $v\geq 2\phi(m_1,m_2,\ldots,m_{n})+1$ where $\phi(m_1,m_2,\ldots,m_{n})$ is defined in \eqref{eq_sum_f}.
			\item In particular, for any positive integer $q$, $m_1 = \cdots = m_{n-1}=q, m_n = (2^{r+2}-2L-1)q/(2^{r+1}-L)$, we obtain an $(L,K=v,M,N)$ scheme with memory ratio $M/N = 1 -2^r(2q/(1+2q))^n/(2^{r+1}-L)$, subpacketization $F = K = v = (2^{r+2} - 2L - 1)(1 + 2q)^n$, and sum-DoF $g = 2^{n+r}$.
		\end{itemize}
		\item Both theoretical and numerical comparisons show that our proposed scheme slightly sacrifices sum-DoF but achieves a significant reduction in subpacketization for some parameters compared to \cite{YWCC, NPR}. Moreover, the proposed scheme achieves lower subpacketization and larger sum-DoF than \cite{WCC, CTWWL} with linear subpacketization under some system parameters.
	\end{itemize}
	
	\subsection{Organizations and Notations}
	The rest of this paper is organized as follows. In Section~\ref{sec-perlimin}, we review the MISO coded caching system and introduce the definition of MAPDA. In Section~\ref{sec-LHSDP}, we introduce the half-sum disjoint packing (HSDP) and show that it can be used to generate a MAPDA. In Section~\ref{sec-Construct-LHSDP}, we propose a new class of MAPDAs by constructing $L$-HSDPs. In Section~\ref{sec-perf-ana}, we provide the performance analysis of the proposed scheme. Finally, we conclude this paper in Section~\ref{sec-conclu}.
	
	\textbf{Notation:} In this paper, we will use the following notations. Let bold capital letter, bold lowercase letter, and curlicue letter denote array, vector, and set respectively; let $|A|$ denote the cardinality of the set $A$; define $[a] = \{1, 2, \ldots, a\}$ and $[a : b]$ is the set $\{a, a+1, \ldots, b-1, b\}$; $\lfloor a \rfloor$ denotes the largest integer not greater than $a$. $a \nmid b$ indicates that $a$ does not divide $b$. We define that $\langle K \rangle_t = K \bmod t$; $\mathbb{Z}_v$ is the ring of integer residues modulo $v$. For any array $\mathbf{P}$, $\mathbf{P}(i,j)$ denotes the entry in the $i^{th}$ row and $j^{th}$ column.
	
	\section{PRELIMINARIES}\label{sec-perlimin}
	In this section, we review the MISO coded caching system, the concept of MAPDA, and their relationship.
	\subsection{System Model}\label{system model}
\begin{figure}[h]
		\centering
		\includegraphics[height=6cm,width=10cm]{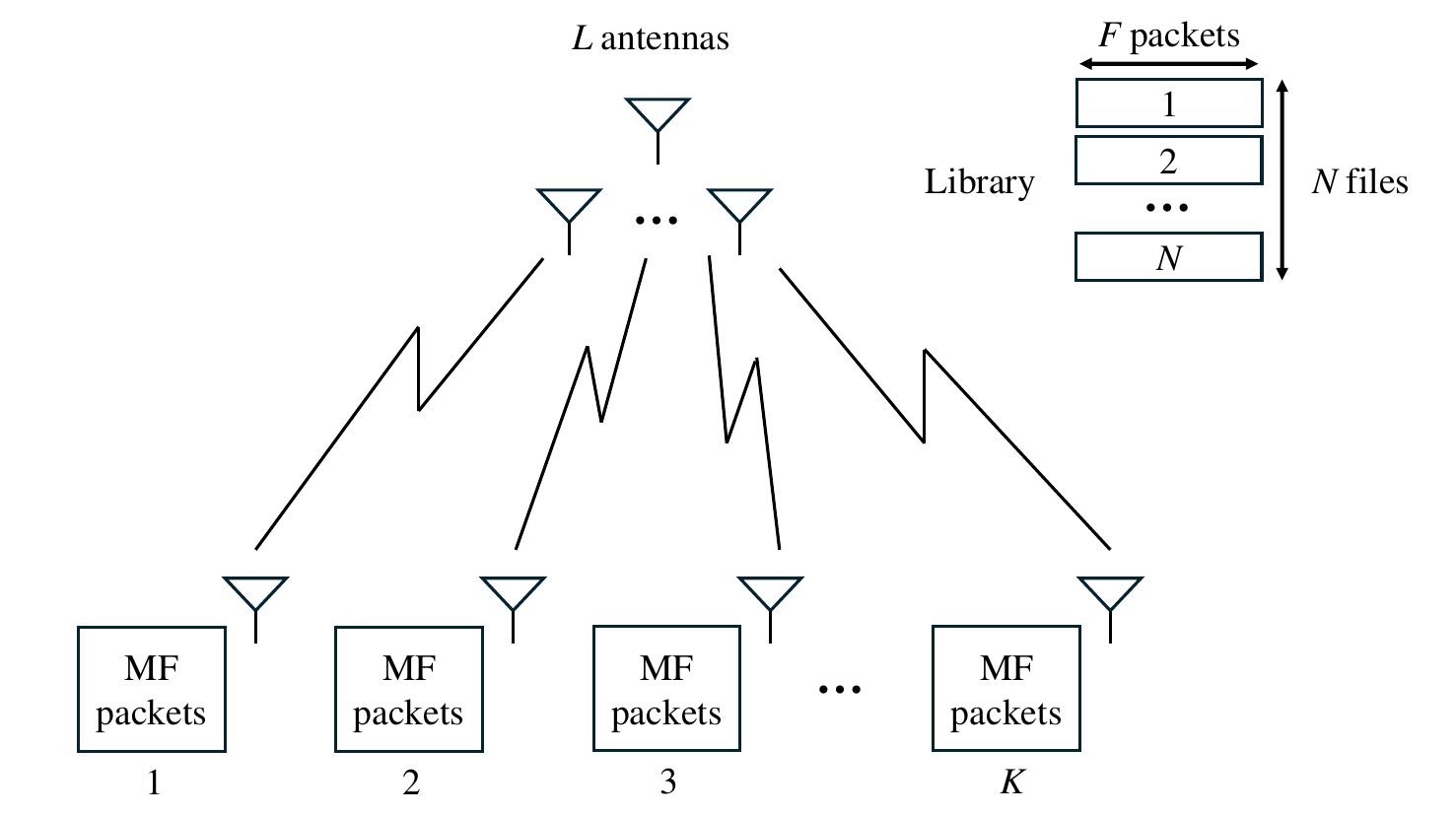}
		\caption{$(L,K,M,N)$ MISO BC system}
		\label{model_fig}
	\end{figure}

	We consider an $(L,K,M,N)$ cache-aided multiple-input single-output (MISO) broadcast channel (BC) system which contains a single server and $K$ users, as illustrated in Fig.~\ref{model_fig}. The server has access to a library of $N$ files, denoted by $\mathcal{W}=\{\mathbf{w}_n \mid n\in[N]\}$, where $N\le K$. Each user is equipped with a cache of size $M$ files where $0\le M\le N$. A server equipped with $L$ antennas communicates with $K$ single-antenna users over the following  wireless broadcast channel.
	At any time slot $t$, the signal transmitted by antenna $i\in[L]$ is denoted by $X_i(t)\in\mathbb{C}$, and satisfies the power constraint $\mathbb{E}[\sum_{i=1}^{L}|X_i(t)|^2]\le P$. The signal received by user $k$ at slot $t$ is
	\begin{align*}
		Y_k(s)=\sum_{i=1}^{L} h_{k,i} X_i(t)+\epsilon_k(t),
	\end{align*}
	where $\epsilon_k(t)\sim\mathcal{CN}(0,1)$ denotes the receiver noise, and $h_{k,i}\in\mathbb{C}$ chosen from independent and identically distributed is the channel gain between antenna $i$ and user $k$. All channel coefficients form the $K\times L$ channel matrix
	\begin{align*}
		\mathbf{H}=
		\begin{pmatrix}
			h_{1,1} & h_{1,2} & \cdots & h_{1,L}\\
			h_{2,1} & h_{2,2} & \cdots & h_{2,L}\\
			\vdots  & \vdots  & \ddots & \vdots \\
			h_{K,1} & h_{K,2} & \cdots & h_{K,L}
		\end{pmatrix},
	\end{align*}
	which is assumed to remain constant during the entire communication process and to be perfectly known to both the server and all users. Based on the property of i.i.d, any subsquare of $\mathbf{H}$ is invertible. An $F$-division $(L,K,M,N)$ MISO coded caching scheme consists of two phases.
	\begin{itemize}
		\item \textbf{Placement phase:}
		Each file $\mathbf{w}_n$ is partitioned into $F$ equal-size packets, i.e., $\mathbf{w}_n=\{\mathbf{w}_{n,f}\mid f\in[F]\}$, where each packet consists of $B$ i.i.d. uniformly distributed bits. The server places a subset of packets into each user's cache. The packets cached by user $k$ is denoted by $\mathcal{Z}_k$ which has at most $MF$ packets. This phase takes place without knowledge of future user demands. 
		
		\item \textbf{Delivery phase:}
		Each user $k\in[K]$ randomly requests a file $\mathbf{w}_{d_k}$ from the library, forming the demand vector $\mathbf{d}=(d_1,\ldots,d_K)$ where $d_k\in[N]$. The server encodes each transmitted packet into a coded packet $\tilde{\mathbf{w}}_{n,f}\in\mathbb{C}^{\tilde B}$ using a Gaussian channel code of rate $B/\tilde B = \log P + o(\log P)$. Assume that there are $S$ transmission intervals, each consisting of $\tilde B$ time slots, under one-shot linear delivery strategy. In transmission interval $s$, the server simultaneously serves $r_s$ users
		$\mathcal{K}_s=\{k_1,\ldots,k_{r_s}\}$ by transmitting the coded packets
		\begin{align*}
			\mathcal{W}_s=\{\tilde{\mathbf{w}}_{d_{k_1},f_1},\ldots,\tilde{\mathbf{w}}_{d_{k_{r_s}},f_{r_s}}\}.
		\end{align*}The signal transmitted by antenna $i\in[L]$ is
		\begin{align*}
			\mathbf{x}_i(s)= \sum_{j\in[r_s]} v_{i,k_j}(s)\,\tilde{\mathbf{w}}_{d_{k_j},f_j},
		\end{align*}
		where $v_{i,k_j}(s)\in\mathbb{C}$ is a precoding coefficient.
		Let $\mathbf{v}_j(s)=(v_{1,k_j}(s),\ldots,v_{L,k_j}(s))$. Then the transmitted signal vector across the $L$ antennas is
		\begin{align}
			\label{Xs}
			\mathbf{X}(s) = \begin{pmatrix}
				\mathbf{x}_1(s) \\
				\mathbf{x}_2(s) \\
				\vdots \\
				\mathbf{x}_L(s)
			\end{pmatrix} = 
			\begin{pmatrix}
				v_{1,k_1}(s) &  \cdots & v_{1,k_{r_s}}(s)\\
				v_{2,k_1}(s) &  \cdots & v_{2,k_{r_s}}(s)\\
				\vdots  & \ddots & \vdots \\
				v_{L,k_1}(s) & \cdots & v_{L,k_{r_s}}(s)
			\end{pmatrix}
			\begin{pmatrix}
				\tilde{\mathbf{w}}_{d_{k_1}, f_1} \\
				\tilde{\mathbf{w}}_{d_{k_2}, f_2} \\
				\vdots \\
				\tilde{\mathbf{w}}_{d_{k_{r_s}}, f_{r_s}}
			\end{pmatrix}
			= \big(\mathbf{v}^{\rm T}_{k_1}(s),\mathbf{v}^{\rm T}_{k_2}(s),\cdots,\mathbf{v}^{\rm T}_{k_s}(s)\big)
			\begin{pmatrix}
				\tilde{\mathbf{w}}_{d_{k_1}, f_1} \\
				\tilde{\mathbf{w}}_{d_{k_2}, f_2} \\
				\vdots \\
				\tilde{\mathbf{w}}_{d_{k_{r_s}}, f_{r_s}}
			\end{pmatrix}.
		\end{align}
		Through the wireless broadcast channel, user $k$ receives the signal
		\begin{align*}
			\mathbf{y}_k(s)= \sum_{i=1}^{L} h_{k,i}\mathbf{x}_i(s) + \epsilon_k(s).
		\end{align*}
		Collecting the received signals of the $r_s$ served users in transmission interval $s$, we obtain
		\begin{equation}
			\label{Ys}
			\begin{aligned}
				\mathbf{Y}(s)&=
				\begin{pmatrix}
					\mathbf{y}_{k_1}(s) \\
					\mathbf{y}_{k_2}(s) \\
					\vdots \\
					\mathbf{y}_{k_{r_s}}(s)
				\end{pmatrix}
				=
				\begin{pmatrix}
					h_{k_1,1} & \cdots & h_{k_1,L}\\
					h_{k_2,1} & \cdots & h_{k_2,L}\\
					\vdots  & \ddots & \vdots \\
					h_{k_{r_s},1} & \cdots & h_{k_{r_s},L}
				\end{pmatrix}
				\begin{pmatrix}
					v_{1,k_1}(s) &  \cdots & v_{1,k_{r_s}}(s)\\
					v_{2,k_1}(s) &  \cdots & v_{2,k_{r_s}}(s)\\
					\vdots  &  \ddots & \vdots \\
					v_{L,k_1}(s) & \cdots & v_{L,k_{r_s}}(s)
				\end{pmatrix}
				\begin{pmatrix}
					\tilde{\mathbf{w}}_{d_{k_1}, f_1} \\
					\tilde{\mathbf{w}}_{d_{k_2}, f_2} \\
					\vdots \\
					\tilde{\mathbf{w}}_{d_{k_{r_s}}, f_{r_s}}
				\end{pmatrix} +
				\begin{pmatrix}
					\epsilon_{k_1}(s) \\
					\epsilon_{k_2}(s) \\
					\vdots \\
					\epsilon_{k_{r_s}}(s)
				\end{pmatrix}\\
				&=
				\mathbf{H}(\mathcal{K}_s, [L])
				\big(\mathbf{v}^{\rm T}_{k_1}(s),\mathbf{v}^{\rm T}_{k_2}(s),\cdots,\mathbf{v}^{\rm T}_{k_s}(s)\big)
				\begin{pmatrix}
					\tilde{\mathbf{w}}_{d_{k_1}, f_1} \\
					\tilde{\mathbf{w}}_{d_{k_2}, f_2} \\
					\vdots \\
					\tilde{\mathbf{w}}_{d_{k_{r_s}}, f_{r_s}}
				\end{pmatrix}
				+
				\begin{pmatrix}
					\epsilon_{k_1}(s) \\
					\epsilon_{k_2}(s) \\
					\vdots \\
					\epsilon_{k_{r_s}}(s)
				\end{pmatrix},
			\end{aligned}
		\end{equation}
		where $\mathbf{H}(\mathcal{K}_s, [L])$ denotes the submatrix of $\mathbf{H}$ with row indices $\mathcal{K}_s$ and all $L$ columns. For each user $k\in\mathcal{K}_s$, all cache-stored interference packets in $\mathbf{y}_k(s)$ can be removed using its cache contents. The remaining interference is eliminated via zero-forcing precoding by properly designing the precoding vectors $\mathbf{v}_j(s)$. After interference cancellation, user $k$ effectively obtains
		\begin{align}
			\mathcal{L}_{s,k}\big(\mathbf{y}_k(s),\tilde{\mathcal{Z}}_k\big)
			= \tilde{\mathbf{w}}_{d_k,f_k} + \varepsilon_k(s),
		\end{align}
		which is equivalent to a point-to-point Gaussian channel at rate $\log P + o(\log P)$. Hence, for sufficiently large $P$, i.e., at high Signal-to-Noise Ratio (SNR), the noise can be ignored and user $k$ can reliably decode its desired coded packet $\mathbf{w}_{d_k,f_k}$.
	\end{itemize}
	
	In each transmission interval $s \in [S]$, the server performs a one-shot linear transmission that simultaneously serves $r_s$ users. Each of these $r_s$ users can successfully decode one desired coded packet from the received signal. Since each coded packet corresponds to one Degree-of-Freedom (DoF), the overall sum-DoF of the system is defined as $\sum_{s=1}^{S} r_s/S$. The objective is to design a scheme that maximizes the achievable sum-DoF for all possible worst-case demands while keeping the subpacketization as low as possible.

	\subsection{Multiple-Antenna Placement Delivery Array}
	The author in \cite{YWCC,NPR} proposed a combinatorial structure called multiple-antenna placement delivery array (MAPDA)  which characterizes the placement and delivery issue in MISO coded caching system. Its definition is given as follows.
	\begin{definition}[MAPDA\cite{YWCC,NPR}]\label{def-MAPDA}\rm
		For positive integers $L$, $K$, $F$, $Z$ and $S$, an $F\times K$ array $\mathbf{P}$ composed of ``$*$'' and $[S]$, is called an $(L, K, F, Z, S)$ multiple-antenna placement delivery array (MAPDA) if it satisfies the following conditions
		\begin{enumerate}
			\item [C$1$.] Each column has exactly $Z$ stars.
			\item [C$2$.] Each integer in $[S]$ occurs at least once.
			\item [C$3$.] Each integer appears at most once in each column;
			\item [C$4$.] For any integer $s\in[S]$, define $\mathbf{P}^{(s)}$ as the subarray of $\mathbf{P}$ including the rows and columns containing $s$ and $r_s'\times r_s$ as the dimension of $\mathbf{P}^{(s)}$. The number of integer entries in each row of $\mathbf{P}^{(s)}$ is no more than $L$, i.e.,
			\begin{align*}
				\left|\left\{k_1\in[r_s]\mid\mathbf{P}^{(s)}(f_1,k_1)\in[S]\right\}\right|\leq L,\ \forall f_1\in[r_s'].
			\end{align*}
		\end{enumerate}
	\end{definition}
	Let us take the following example to further explain the concept of MAPDA. 
	
	\begin{example}\rm
		\label{ex-MAPDA}
		The following array is a $(L=3,K=4,F=4,Z=1,S=3)$ MAPDA.
		\begin{align*}
			\mathbf{P} =
			\begin{blockarray}{ccccc}
				1 & 2 & 3 & 4 \\
				\begin{block}{(cccc)c}
					* & 1 & 2 & 3 & 1 \\
					1 & * & 3 & 2 & 2 \\
					2 & 3 & * & 1 & 3 \\
					3 & 2 & 1 & * & 4 & \\
				\end{block}
			\end{blockarray}. 
		\end{align*}  
		From $\mathbf{P}$, it can be seen that each column contains exactly $Z=1$ star, and each integer appears exactly once in every column. Therefore, conditions C$1$, C$2$, and C$3$ of Definition~\ref{def-MAPDA} are satisfied. Next, consider Condition C4. For all three subarrays of $\mathbf{P}$ corresponding to the integers $1,2,3$, we have
		\begin{align*}
			\mathbf{P}^{(1)}=\mathbf{P}^{(2)}=\mathbf{P}^{(3)}=
			\begin{pmatrix}
				* & 1 & 2 & 3 \\
				1 & * & 3 & 2 \\
				2 & 3 & * & 1 \\
				3 & 2 & 1 & *
			\end{pmatrix}.
		\end{align*}
		Each row of these subarrays contains at most $L=3$ integers, and hence condition C$4$ is also satisfied.
		
		{\em By interpreting the rows and columns of $\mathbf{P}$ as subpacketization and users respectively, letting a star represent the user’s caching status, and letting each integer indicate both the transmission interval and the delivery strategy,} we can obtain the following $(L=3,K=4,M,N)$ MISO coded caching scheme with memory size $M=N/4$ files.
		\begin{itemize}
			\item \textbf{Placement phase:} Each file $\mathbf{w}_{n}$ is divided into $F=4$ packets denoted as $\mathbf{w}_{n}=\{\mathbf{w}_{n,1},\mathbf{w}_{n,2},\mathbf{w}_{n,3},\mathbf{w}_{n,4}\}$. For any $f\in[F]=[4]$ and $k\in[K]=[4]$, $\mathbf{P}(f,k)=*$ represents that user $k$ caches the $f^{th}$ packet of all files, then the cached content of all users are
			\begin{align*}
				\mathcal{Z}_{1} =\{\mathbf{w}_{n,1}|n\in[N]\},\ \mathcal{Z}_{2}=\{\mathbf{w}_{n,2}|n\in[N]\}, \  
				\mathcal{Z}_{3} =\{\mathbf{w}_{n,3}|n\in[N]\},\ \mathcal{Z}_{4}=\{\mathbf{w}_{n,4}|n\in[N]\}. 
			\end{align*} 
			Then each user caches exactly $N \cdot Z= N$ packets, i.e., $M=\frac{N}{4}$ file.
			
			\item \textbf{Delivery phase:} Assume that the demand vector is $\mathbf{d}=(1,2,3,4)$. If the integer entry $\mathbf{P}(f,k)=s$, the user $k$ does not cache the required packet $\mathbf{w}_{d_k,f}$. For instance, the entries $\mathbf{P}(2,1)=\mathbf{P}(1,2)=\mathbf{P}(4,3)=\mathbf{P}(3,4)=1$ imply that users $1$, $2$, $3$ and $4$ require the packets $\mathbf{w}_{1,2}$, $\mathbf{w}_{2,1}$, $\mathbf{w}_{3,4}$ and $\mathbf{w}_{4,3}$, respectively. Then the server encodes these packets into  $\hat{\mathcal{W}}_1=\{\tilde{\mathbf{w}}_{1,2},\tilde{\mathbf{w}}_{2,1},\tilde{\mathbf{w}}_{3,4}, \tilde{\mathbf{w}}_{4,3}\}$. Assume that the channel matrix
			\begin{align*}
				\mathbf{H}=
				\begin{pmatrix}
					1 & 2 & 4\\
					1 & 3 & 9\\
					1 & 4 & 16\\
					1 & 5 & 25
				\end{pmatrix}.
			\end{align*}
			
			Note that $\tilde{\mathbf{w}}_{2,1},\tilde{\mathbf{w}}_{1,2}$ are the interferences of users $3,4$ and $\tilde{\mathbf{w}}_{4,3},\tilde{\mathbf{w}}_{3,4}$ are the interferences of users $1,2$, since these packets are neither cached nor requested by the corresponding users. To eliminate the interfering packets that are neither cached nor requested by the corresponding users, we design the precoding column vectors $\mathbf{v}_{1}^{\rm T}(1),\mathbf{v}_{2}^{\rm T}(1),\mathbf{v}_{3}^{\rm T}(1),\mathbf{v}_{4}^{\rm T}(1)$ so that all these interfering packets are zero-forced at the corresponding users. More precisely, the precoding column vectors are chosen to satisfy
			\begin{equation}
				\label{ZF}
				\begin{aligned}
					\mathbf{h}_1 \mathbf{v}^{\rm T}_3(1) = \mathbf{h}_1 \mathbf{v}^{\rm T}_4(1) &= 0, \quad
					\mathbf{h}_2 \mathbf{v}^{\rm T}_3(1) = \mathbf{h}_2 \mathbf{v}^{\rm T}_4(1) = 0,\\
					\mathbf{h}_3 \mathbf{v}^{\rm T}_1(1) = \mathbf{h}_3 \mathbf{v}^{\rm T}_2(1) &= 0, \quad
					\mathbf{h}_4 \mathbf{v}^{\rm T}_1(1) = \mathbf{h}_4 \mathbf{v}^{\rm T}_2(1) = 0,
				\end{aligned}
			\end{equation}
			where $\mathbf{h}_k=(h_{k,1},h_{k,2},h_{k,3})$ denotes the $k^{th}$ row of the channel matrix 
			$\mathbf{H}$ for each $k\in[4]$. As proved in~\cite{YWCC}, such precoding column vectors always exist. By solving the equations in \eqref{ZF}, we can set  
			\begin{align*}
				\mathbf{v}^{\rm T}_1(1) = \mathbf{v}^{\rm T}_2(1) = \begin{pmatrix}20\\ -9\\ 1\end{pmatrix}, \qquad
				\mathbf{v}^{\rm T}_3(1) = \mathbf{v}^{\rm T}_4(1) = \begin{pmatrix}6\\ -5\\ 1\end{pmatrix}.
			\end{align*}
			Then from~\eqref{Xs}, the transmitted signal vector by the $L=3$ antennas in interval $1$ can be written as
			\begin{align*}
				\mathbf{X}(1) = \begin{pmatrix}
					\mathbf{x}_1(1) \\
					\mathbf{x}_2(1) \\
					\mathbf{x}_3(1)
				\end{pmatrix}
				= \big(\mathbf{v}^{\rm T}_1(1),\mathbf{v}^{\rm T}_2(1),\mathbf{v}^{\rm T}_3(1),\mathbf{v}^{\rm T}_4(1)\big)
				\begin{pmatrix}
					\tilde{\mathbf{w}}_{1, 2} \\
					\tilde{\mathbf{w}}_{2, 1} \\
					\tilde{\mathbf{w}}_{3, 4} \\
					\tilde{\mathbf{w}}_{4, 3}
				\end{pmatrix} = 
				\begin{pmatrix}
					20\tilde{\mathbf{w}}_{1, 2}+20\tilde{\mathbf{w}}_{2, 1}+6\tilde{\mathbf{w}}_{3, 4}+6\tilde{\mathbf{w}}_{4, 3} \\
					-9\tilde{\mathbf{w}}_{1, 2}-9\tilde{\mathbf{w}}_{2, 1}-5\tilde{\mathbf{w}}_{3, 4}-5\tilde{\mathbf{w}}_{4, 3} \\
					\tilde{\mathbf{w}}_{1, 2}+\tilde{\mathbf{w}}_{2, 1}+\tilde{\mathbf{w}}_{3, 4}+\tilde{\mathbf{w}}_{4, 3}
				\end{pmatrix}.
			\end{align*}
			From~\eqref{Ys}, by collecting the received signals of the served users 
			$\mathcal{K}_1=\{1,2,3,4\}$ in transmission interval $1$, we obtain
			\begin{align*}
				\mathbf{Y}(1)=
				\begin{pmatrix}
					\mathbf{y}_{1}(1) \\
					\mathbf{y}_{2}(1) \\
					\mathbf{y}_{3}(1) \\
					\mathbf{y}_{4}(1)
				\end{pmatrix}=
				\mathbf{H}(\mathcal{K}_1, [3])\mathbf{X}(1)
				+
				\begin{pmatrix}
					\epsilon_{1}(1) \\
					\epsilon_{2}(1) \\
					\epsilon_{3}(1) \\
					\epsilon_{4}(1)
				\end{pmatrix}
				=
				\begin{pmatrix}
					6\tilde{\mathbf{w}}_{1, 2}+6\tilde{\mathbf{w}}_{2, 1}\\
					2\tilde{\mathbf{w}}_{1, 2}+2\tilde{\mathbf{w}}_{2, 1}\\
					2\tilde{\mathbf{w}}_{3, 4}+2\tilde{\mathbf{w}}_{4, 3}\\
					6\tilde{\mathbf{w}}_{3, 4}+6\tilde{\mathbf{w}}_{4, 3}
				\end{pmatrix}+
				\begin{pmatrix}
					\epsilon_{1}(1) \\
					\epsilon_{2}(1) \\
					\epsilon_{3}(1) \\
					\epsilon_{4}(1)
				\end{pmatrix}.
			\end{align*}
			Then user $1$ can receive $\mathbf{y}_1(1)= 6\tilde{\mathbf{w}}_{1,2} + 6\tilde{\mathbf{w}}_{2,1} + \epsilon_1(1)$. For sufficiently large $P$ (i.e., high SNR), the noise can be neglected. Since user $1$ has cached $\mathbf{w}_{2,1}$, it can obtain $\tilde{\mathbf{w}}_{2,1}$ and thus decode its desired packet $\tilde{\mathbf{w}}_{1,2}$. Similarly, other users can also decode their requested packets.
		\end{itemize}
		
		In summary, the delivery process consists of $S=3$ transmission intervals since there are $S=3$ distinct integers in $\mathbf{P}$. In each transmission interval, the server transmits $4$ packets to $4$ users, achieving a sum-DoF of $4$.
		\hfill $\square$
	\end{example}
	By the aforementioned example and its explanation, the following result can be obtained.
	\begin{lemma}[{\cite{YWCC}}] \rm
		\label{lemma}
		For a given $(L,K,F,Z,S)$ MAPDA, there exists an $F$-division scheme for the $(L,K,M,N)$ MISO BC system with memory ratio $M/N=Z/F$, sum-DoF $K(F-Z)/S$, and subpacketization $F$.
	\end{lemma}

	\section{HALF-SUM DISJOINT PACKING}
	\label{sec-LHSDP}
	In this section, we first introduce a novel combinatorial structure, termed half-sum disjoint packing (HSDP), which generalize the concept of non-half-sum disjoint packing (NHSDP) in \cite{CWWC} to construct PDA, to construct MAPDA. 
	
	Now, let us see the definition of HSDP as follows.
	\begin{definition}[HSDP]\label{def-LHSDP}\rm
		For any positive integer $L$ and positive odd integer $v$, a pair $(\mathbb{Z}_v,\mathfrak{D})$ where $\mathfrak{D}$ consists of $b$ $g$-subsets of $\mathbb{Z}_v$ denoted as $\mathfrak{D} = \{\mathcal{D}_i = \{d_{i,1}, d_{i,2}, \ldots, d_{i,g}\} | i \in [b]\}$ is called $L$-$(v,g,b)$ half-sum disjoint packing (HSDP) if it satisfies the following conditions.
		\begin{itemize}
			\item  Block-disjoint: The intersection of any two different elements in $\mathfrak{D}$ is empty;  
			\item  L-half-sum:
			For  any $i \in [b]$ and $j \in [g]$, $\sum_{i'=1}^{b}|\mathcal{D}_{i'} \cap \mathcal{B}_{i,j}| < L$ always holds, where 
			\begin{align}
				\label{eq-HS}
				\mathcal{B}_{i, j} = \left\{\frac{d_{i,j} + d_{i,j'}}{2}\ \Big| j' \in [g]\setminus \{j\} \right\}.
			\end{align}
		\end{itemize} \hfill $\square$
	\end{definition}
	It is worth noting that a $1$-HSDP is exactly a NHSDP which is defined in \cite{CWWC}. Let us take the following example to further explain the concept of $L$-HSDP. 
	\begin{example}\rm
		\label{example-by-defination}When $L=2$, $(v,g,b) = (11,4,2)$, let us consider the following $(\mathbb{Z}_{11},\mathfrak{D})$ where  
		\begin{align*}
			\mathfrak{D}= \{\mathcal{D}_1=\{1, 2, 4, 10\},\ \mathcal{D}_2=\{5, 6, 8, 9\}.
		\end{align*}Clearly $\mathcal{D}_1\cap\mathcal{D}_2=\emptyset$, i.e., the first condition of Definition~\ref{def-LHSDP} holds. 
		When $i=1$ and $j=1,2,3,4$ in Definition \ref{def-LHSDP}, from \eqref{eq-HS} we have 
		\begin{align*}
			&\mathcal{B}_{1,1} =\left\{\frac{1+2}{2}, \frac{1+4}{2}, \frac{1+10}{2}\right\} = \{0,7,8\}, \, \mathcal{B}_{1,2}= \left\{\frac{2+1}{2}, \frac{2+4}{2}, \frac{2+10}{2}\right\} = \{3,6,7\},\\
			&\mathcal{B}_{1,3}=\left\{\frac{4+1}{2}, \frac{4+2}{2}, \frac{4+10}{2}\right\} = \{3,7,8\}, \, \mathcal{B}_{1,4}=\left\{\frac{10+1}{2}, \frac{10+2}{2}, \frac{10+4}{2}\right\} = \{0,6,7\}.
		\end{align*}Clearly, $|\{0,7,8\}\cap(\mathcal{D}_1 \cup \mathcal{D}_2)|=|\{8\}| = 1 < L$. Similarly, we can show that the intersection between the set $\mathcal{B}_{i,j}$ and the union of all blocks is strictly smaller than $L=2$. This implies that the second condition of Definition~\ref{def-LHSDP} holds. So $(\mathbb{Z}_{11},\mathfrak{D})$ is a $2$-$(11,4,2)$ HSDP.
		\hfill $\square$
	\end{example}
	
	Having introduced the definition of HSDP and illustrated it through an explicit example, we now turn to explaining how such a combinatorial structure can be transformed into a MAPDA. 
	
	By exploiting the structure of cyclic Latin squares, a MAPDA with linearsubpacketization can be constructed via an $L$-HSDP. The main idea is as follows. Given an $L$-$(v,g,b)$ HSDP, we first construct a Latin square $\mathbf{L}$ of order $v$ by cyclic shifts, which guarantees that every integer occurs at most once in each row and each column. This step ensures conditions C$2$ and C$3$ of Definition~\ref{def-MAPDA}.
	Next, we partition all entries of $\mathbf{L}$ into mutually disjoint orbits, where each orbit is generated by the integer appearing in the first row of $\mathbf{L}$ and has size $v$. We then retain all orbits generated by the integers contained in the blocks of the $L$-HSDP, while replacing the entries in the remaining orbits with the symbol ``$*$''. This step ensures condition C$1$ of Definition~\ref{def-MAPDA}$,$ and moreover guarantees that the entries lying on the orbits generated by the integers in the same block form subarrays that satisfy condition C$4$ of Definition~\ref{def-MAPDA}.
	Finally, to ensure that the entire array satisfies condition C$4$, we further distinguish orbit classes according to their corresponding blocks.
	We now present the following example to illustrate the idea.
	\begin{example}\rm
		\label{example-MAPDA-by-LHSDP}
		We illustrate the procedure using the $2$-$(v=11,g=4,b=2)$ HSDP from Example~\ref{example-by-defination} with $\mathfrak{D}=\{\mathcal{D}_1=\{1, 2, 4, 10\},\ \mathcal{D}_2=\{5, 6, 8, 9\}\}$. The overall transformation from an $L$-HSDP to a MAPDA (see Fig.~\ref{flow-diagram}) consists of the following three steps:
		\begin{figure}[htbp!]
			\centering
			\begin{minipage}{\textwidth}
				\centering
				\includegraphics[width=0.8\textwidth]{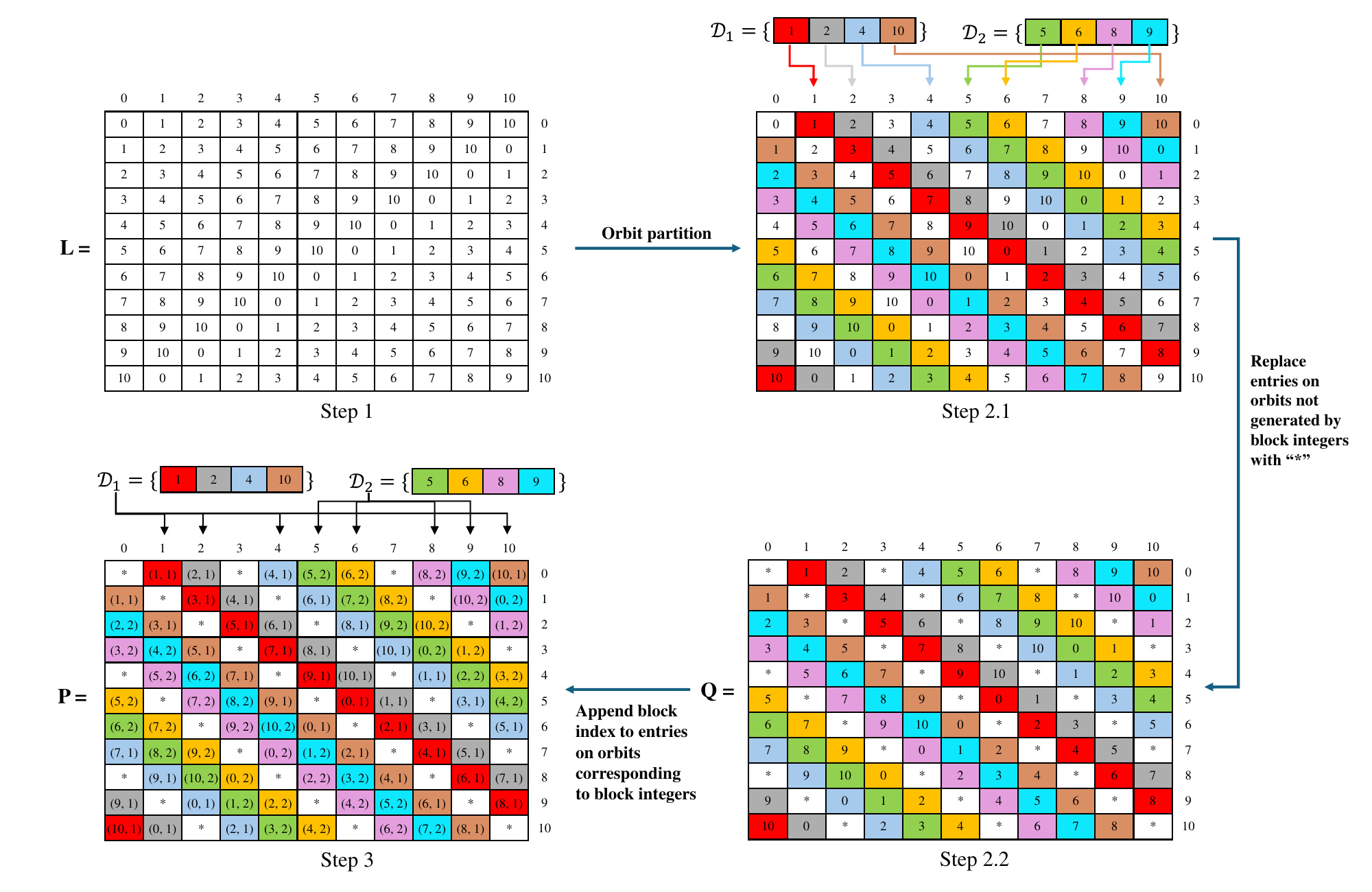}
				\caption{The flow diagram of constructing the array $\mathbf{P}$ via a $2$-$(11,4,2)$ HSDP}
				\label{flow-diagram}
			\end{minipage}
		\end{figure}
		\begin{itemize}
			\item \textbf{Step 1. Latin square construction.}
			We first construct the $11\times 11$ Latin square $\mathbf{L}=(l_{f,k})_{f,k\in \mathbb{Z}_{11}}$ by cyclic shifts, where each entry is given by $l_{f,k}=f+k$. As illustrated in Step $1$ of Fig.~\ref{flow-diagram}, we verify that every integer appears at most once in each row and each 
			column, which ensures that conditions C$2$ and C$3$ of Definition~\ref{def-MAPDA} holds.
			
			\item \textbf{Step 2. Orbit partition.} We partition all entries of $\mathbf{L}$ into mutually disjoint orbits, each of size $v$. 
			Specifically, for each $j\in\mathbb{Z}_{11}$, the orbit generated by the integer $j$ is defined as
			\begin{align*}
				\text{Orbit-}j
				:= \{(f,k)\mid \langle k-f\rangle_{11}=j,\ f,k\in\mathbb{Z}_{11}\}.
			\end{align*}
			After the orbit partition, we process the orbits in the following two substeps.
			
			- \textbf{Step 2.1. Orbit selection.}
			We retain all orbits generated by the integers contained in $\mathcal{D}_1$ and $\mathcal{D}_2$. As illustrated in Step $2.1$ of Fig.~\ref{flow-diagram}, the orbits generated by the integers in $\mathcal{D}_1$ and $\mathcal{D}_2$	are highlighted using different colors. 
			More precisely, Orbit-$1$, Orbit-$2$, Orbit-$4$, Orbit-$5$, Orbit-$6$, Orbit-$8$, Orbit-$9$ and Orbit-$10$ correspond to the red, gray, blue, green, orange, purple, cyan and brown orbits, respectively.
			
			- \textbf{Step 2.2. Star replacement.}
			All entries lying on orbits, each of which is not generated by the integers contained in 
			$\mathcal{D}_1$ and $\mathcal{D}_2$, are replaced by the symbol ``$*$'', yielding the array 
			$\mathbf{Q}$. As illustrated in Step $2.2$ of Fig.~\ref{flow-diagram}, all entries on 
			Orbit-$0$, Orbit-$3$, and Orbit-$7$ are replaced by ``$*$''. Then, each column of 
			$\mathbf{Q}$ contains exactly $Z = v - bg = 11 - 2 \times 4 = 3$ stars, which ensures 
			condition C$1$ of Definition~\ref{def-MAPDA}.
			Moreover, for any given block, the subarray formed by all entries lying on the orbits generated by the integers in that block satisfies condition C$4$ Definition~\ref{def-MAPDA}.
			For instance, we consider the subarray of $\mathbf{Q}$ induced by the rows and columns containing the entry $s=1$ on Orbit-$1$, Orbit-$2$, Orbit-$4$, and Orbit-$10$, which are generated by the integers in $\mathcal{D}_1$, and the subarray $\mathbf{Q}^{(1)}_{\text{Orbit-}\{1,2,4,10\}}$ is shown in Fig.~\ref{subarray}.
			\begin{figure}[htbp!]
				\centering
				\begin{minipage}{\textwidth}
					\centering
					\includegraphics[width=0.7\textwidth]{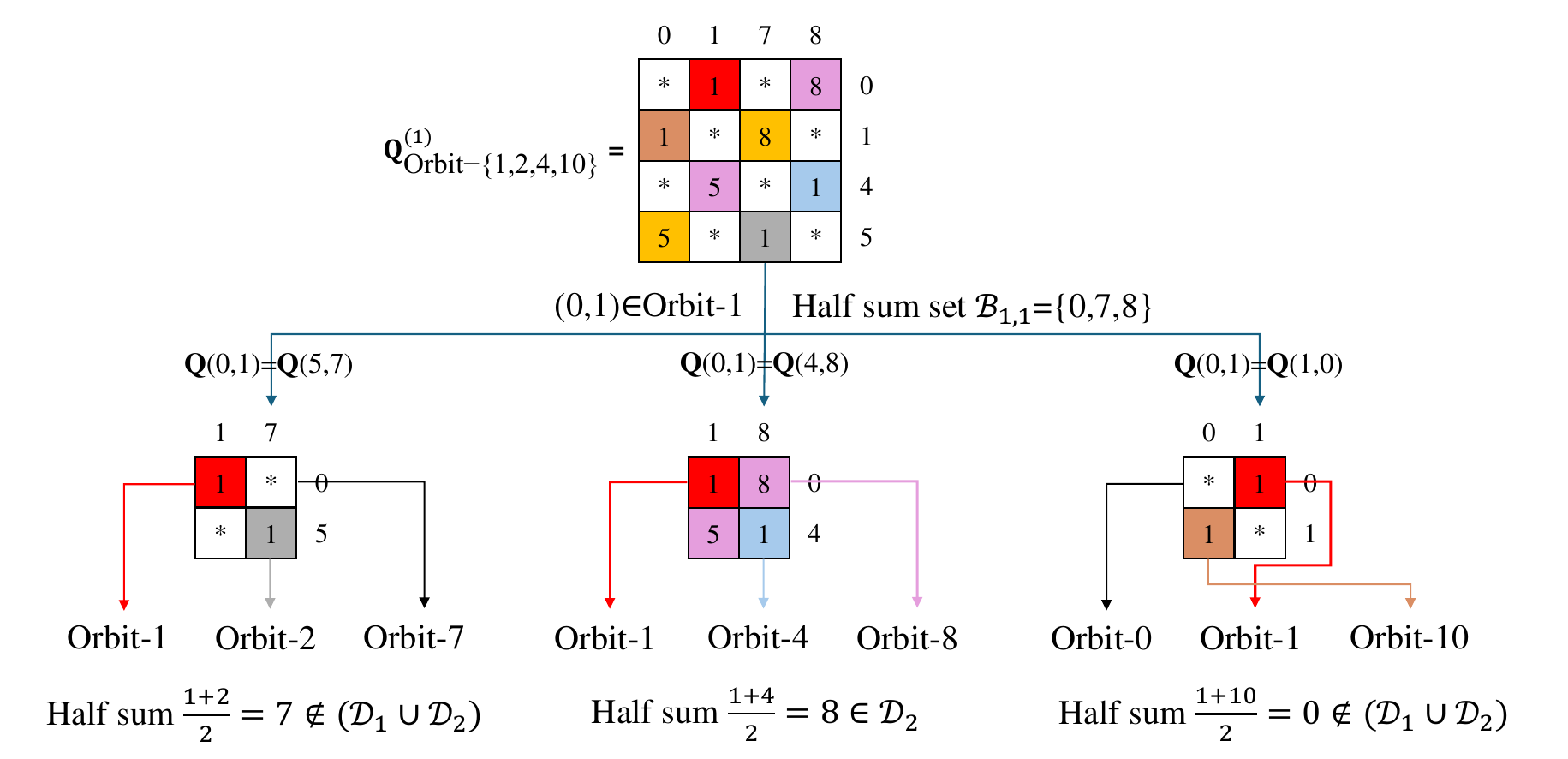}
					\caption{The subarray $\mathbf{Q}^{(1)}_{\text{Orbit-}\{1,2,4,10\}}$: Let us see the first row containing the entry $\mathbf{Q}(0,1)$.
						Since $(0,1) \in \text{Orbit-}1$ which is generated by the integer $1 \in \mathcal{D}_1$, the corresponding half-sum set is $\mathcal{B}_{1,1} = \{0,7,8\}$.
						We first consider the pair $\mathbf{Q}(0,1) = \mathbf{Q}(5,7)$, where $(5,7) \in \text{Orbit-}2$.
						Since the half-sum $(1+2)/2 = 7 \notin (\mathcal{D}_1 \cup \mathcal{D}_2)$, it follows that $\mathbf{Q}(0,7) = *$, where $(0,7) \in \text{Orbit-}7$.
						Next, consider the pair $\mathbf{Q}(0,1) = \mathbf{Q}(4,8)$, where $(4,8) \in \text{Orbit-}4$.
						Since the half-sum $(1+4)/2 = 8 \in \mathcal{D}_2$, we have $\mathbf{Q}(0,8) = 8$, where $(0,8) \in \text{Orbit-}8$.
						Finally, consider the pair $\mathbf{Q}(0,1) = \mathbf{Q}(1,0)$, where $(1,0) \in \text{Orbit-}10$.
						Since the half-sum $(1+10)/2 = 0 \notin (\mathcal{D}_1 \cup \mathcal{D}_2)$, it follows that $\mathbf{Q}(0,0) = *$, where $(0,0) \in \text{Orbit-}0$. Clearly, $\bigl|\mathcal{B}_{1,1} \cap (\mathcal{D}_1 \cup \mathcal{D}_2)\bigr|= |\{8\}| = 1 < L = 2$ which implies that the first row has at most $L=2$ non-$*$ entries.}
					\label{subarray}
				\end{minipage}
			\end{figure} 
			It is not difficult to check that each row of $\mathbf{Q}^{(1)}_{\text{Orbit-}\{1,2,4,10\}}$
			contains at most $L=2$ integer entries, and hence condition C$4$ is satisfied. 
			By the same argument, the number of non-$*$ entries in each rows of $\mathbf{Q}^{(1)}_{\text{Orbit-}\{1,2,4,10\}}$ can be verified in an analogous manner.
			Hence, the subarray $\mathbf{Q}^{(1)}_{\text{Orbit-}\{1,2,4,10\}}$ satisfies condition C$4$ of Definition~\ref{def-MAPDA}.
			It is worth noting that the array $\mathbf{Q}$ does not satisfy condition C$4$. 
			This is because identical integer entries may appear on orbits associated with different blocks.
			For instance, consider the subarray $\mathbf{Q}^{(1)}$, given by
			\begin{align*}
				\mathbf{Q}^{(1)} =
				\begin{blockarray}{ccccccccc}
					0 & 1 & 3 & 5 & 7 & 8 & 9 & 10 \\
					\begin{block}{(cccccccc)c}
						* & 1 & * & 5 & * & 8 & 9 & 10 & 0\\
						1 & * & 4 & 6 & 8 & * & 10 & 0 & 1\\
						2 & 3 & 5 & * & 9 & 10 & * & 1 & 2\\
						3 & 4 & * & 8 & 10 & 0 & 1 & * & 3\\
						* & 5 & 7 & 9 & * & 1 & 2 & 3 & 4\\
						5 & * & 8 & * & 1 & * & 3 & 4 & 5\\
						7 & 8 & * & 1 & * & 4 & 5 & * & 7\\
						9 & * & 1 & * & 5 & 6 & * & 8 & 9\\
					\end{block}
				\end{blockarray}.
			\end{align*}
			We observe that the number of integer entries in each row of $\mathbf{Q}^{(1)}$ is greater than $L=2$. So condition C$4$ is violated. 
			
			\item \textbf{Step 3. Block index assignment.}
			Finally, for each orbit generated by the integers contained in $\mathcal{D}_1$ and $\mathcal{D}_2$,
			we assign the corresponding block index as the second coordinate to every entry on that orbit,
			yielding the array $\mathbf{P}$.
			By distinguishing orbit classes according to their associated blocks,
			this step ensures that $\mathbf{P}$ satisfies condition~C$4$ of Definition~\ref{def-MAPDA}.
			As illustrated in Step~3 of Fig.~\ref{flow-diagram},
			entries on Orbit-$1$, Orbit-$2$, Orbit-$4$, and Orbit-$10$ are assigned block index $1$,
			whereas entries on Orbit-$5$, Orbit-$6$, Orbit-$8$, and Orbit-$9$ are assigned block index $2$.
			To verify condition C$4$, consider the subarray $\mathbf{P}^{(1,1)}$, given by
			\begin{align*}
				\mathbf{P}^{(1,1)} =
				\begin{blockarray}{ccccc}
					0 & 1 & 7 & 8 \\
					\begin{block}{(cccc)c}
						* & (1,1) & * & (8,2) & 0\\
						(1,1) & * & (8,2) & * & 1\\
						* & (5,2) & * & (1,1) & 4\\
						(5,2) & * & (1,1) & * & 5\\
					\end{block}
				\end{blockarray}.
			\end{align*}
			Clearly, each row of $\mathbf{P}^{(1,1)}$ contains at most $L=2$ vector entries. Similarly, we can verify that all other vector entries in $\mathbf{P}$ satisfy condition C$4$. Therefore, $\mathbf{P}$ is a $(2,11,11,3,22)$ MAPDA.
		\end{itemize}
		
		By Lemma~\ref{lemma} the $(2,11,11,3,22)$ MAPDA $\mathbf{P}$ can realizes an $F$-division $(L=2,K=11,M,N)$ MISO coded caching scheme with memory ratio $M/N = 3/11$, sum-DoF $g = 4$ and subpacketization $F=K=11$. Now let us consider the existing schemes with $K=11$ and $M/N=3/11$.
		\begin{itemize}
			\item The YWCC $1$ scheme: In \cite{YWCC} we have a $(L = 2, K=11,M,N)$ YWCC $1$ scheme with the memory ratio $M/N = 3/11$, the subpacketization $F_{\text{YWCC}} = 5\binom{11}{3} = 825 > 11=F$, the sum-DoF $g_{\text{YWCC}} = 5 > 4 = g$. So our scheme achieves a significant reduction in subpacketization, while the sum-DoF loss is only 1 compared to YWCC $1$ Scheme.
			
			\item  The CTWWL scheme: In \cite{CTWWL} we have a $(L=2,K=11,M,N)$ CTWWL scheme with the memory ratio $M/N=3/11$, the subpacketization $F_{\text{CTWWL}} = 44 > 11 = F$, the sum-DoF $g_{\text{CTWWL}} = 4 = g$. So our scheme achieves lower subpacketization while maintaining the same sum-DoF and memory ratio.
		\end{itemize}
		\hfill $\square$
	\end{example} 
	
	Based on the above example, we next give the general mathematical expression of the construction.
	\begin{construction}\rm\label{cons-MAPDA via LHSDP}
		Given an $L$-$(v,g,b)$ HSDP $(\mathbb{Z}_v,\mathfrak{D})$, we can construct a $v\times v$ array $\mathbf{P}$ where for each integers $f,k\in\mathbb{Z}_v$, the entry 
		\begin{equation}\label{eq-cons-1}
			\mathbf{P}(f,k)=\begin{cases}
				(f+k,i),& \mbox{if\ } k-f \in \mathcal{D}_i,\ \exists\ i\in[b];\\
				\ \ \ \ \ *, &\mbox{otherwise}.
			\end{cases}
		\end{equation} \hfill $\square$
	\end{construction}
	
	In fact, for any $L$-$(v,g,b)$ HSDP the obtained array generated by Construction \ref{cons-MAPDA via LHSDP} is always a MAPDA with $F=K=v$. That is the following result whose proof is included in Appendix~\ref{proof of MAPDA to LHSDP}. 
	\begin{theorem}[MAPDA via $L$-HSDP]\rm\label{MAPDA via LHSDP}
		Given an $L$-$(v,g,b)$ HSDP, we can obtain an $(L,K=v,F=v,Z=v-bg,S=bv)$ MAPDA which leads to an $(L,K=v,M,N)$ MISO coded caching scheme with memory ratio $M/N=1-bg/v$, sum-DoF $g$, subpacketization $F=v$.\hfill $\square$
	\end{theorem}
	
	By Theorem \ref{MAPDA via LHSDP}, we can generate a MAPDA by constructing an $L$-HSDP. So it is meaningful to construct $L$-HSDP with good performance. 
	
	\section{A General Construction Framework for $L$-HSDPs}
	\label{sec-Construct-LHSDP}
	In this section, we propose a novel construction framework for $L$-HSDPs. In addition, we derive an optimal solution by transforming the parameter selection as an integer optimization problem under this framework.
	
	Unlike the NHSDP construction, the main challenge in our design is to construct a class of $L$-HSDPs such that the intersection between the half-sum set and the union of all blocks is as large as possible (up to $L-1$). Equivalently, minimize the memory ratio under fixed subpacketization and sum-DoF.
	
	Our framework for constructing $L$-HSDPs is proposed based on embedding integers into a high-dimensional geometric space.
	The core idea is to design a subset $\mathcal{X}$ such that both block integers and their corresponding half-sums admit linear representations over $\mathcal{X}$, where the representations of block integers are not necessarily unique.
	Specifically, we construct a basis set $\mathcal{X}$ consisting of $n+r$ ordered elements, where the last $r$ elements (referred to as the ``tail'') are designed to control the size of intersection between blocks and their corresponding half-sum sets, subject to the parameter $L$. Using this basis set $\mathcal{X}$, we form an $L$-HSDP $(\mathbb{Z}_v,\mathfrak{D})$.
	
		%
		%
	
	\begin{construction}\rm
		\label{cons-LHSDP}
		For any given positive integer $L$, there exists a smallest positive integer $r$ such that $L \leq 2^r$. For any $n$ positive integers $m_1$, $m_2$, $\ldots$, $m_{n}$, let $\mathcal{A}:=[m_1]\times [m_2]\times \cdots\times[m_{n}]$ and $m_{n+1} = m_{n+2} = \dots = m_{n+r} = 1$. Define the following recursive function
		\begin{align}\label{recursive-constr}
			f(i)=\begin{cases}
				m_1\ \ \ \ \ \ \ \ \ \ \ \ \ \ \ \ \ \ \ \ \ \ \ \ &\ \text{if}\ i=1,\\
				m_i\left(2\sum_{j=1}^{i-1}f(j)+1\right) &\ \text{if}\ 2 \leq i\leq n,\\
				(m_{i-1}+1)\frac{f(i-1)}{m_{i-1}} &\ \text{if}\ n+1 \leq i \leq n+r-1,\\
				(m_{n+r-1}+1)\frac{f(n+r-1)}{m_{n+r-1}} + (2^r - L)f(n+1) &\ \text{if}\ i=n+r. 
			\end{cases}
		\end{align}
		and $\mathcal{X}:=\{x_i=\frac{f(i)}{m_i} | i\in [n+r]\}$.  We can construct a family $\mathfrak{D}=\{\mathcal{D}_{\bf a}\ |\ {\bf a}=(a_1,a_2,\ldots,a_{n})\in \mathcal{A}\}$ where $\mathcal{D}_{\bf a}$ is defined as  
		\begin{align}
			\label{eq-family}
			\mathcal{D}_{\bf a}=\left\{(\alpha_1 a_1x_1+\cdots+\alpha_n a_nx_n)+(\alpha_{n+1}x_{n+1}+\cdots+\alpha_{n+r}x_{n+r})\ |\  \alpha_i\in\{-1,1\},i\in[n+r]\right\},
		\end{align}for each vector ${\bf a}=(a_1,a_2,\ldots,a_{n})\in \mathcal{A}$.  By the above construction, $\mathfrak{D}$ has $m_1 \times m_2 \times \cdots \times m_{n}$ blocks, and each block $\mathcal{D}_{\bf a}$ has $2^{n+r}$ integers. 
		\hfill $\square$ 
	\end{construction}
	
	Let us take the following example to illustrate Construction~\ref{cons-LHSDP}.
	\begin{example}\rm\label{exam-constr}
		When $L=4$, $r=2$, $n=2$, $m_1=2$ and $m_2=2$, we have $\mathcal{A}=[2]\times[2]$ and $m_{n+1}=m_{n+r}=m_3 = m_4=1$. From \eqref{recursive-constr}, we have 
		
		\begin{align*}
			&f(1)=m_1=2,\ \  f(2)=m_2(2f(1)+1)=2(4+1)=10, \\ 
			&f(3)=(m_2+1)\frac{f(2)}{m_2}=3 \cdot 5=15, \ \ f(4)=(m_3+1)\frac{f(3)}{m_3} + (2^2-2)f(2)=(1+1)15 = 30,
		\end{align*}respectively. Then we have the subset 
		\begin{align*}
			\mathcal{X}=\left\{x_1=\frac{f_1}{m_1}=\frac{2}{2}=1, x_2=\frac{f(2)}{m_2}=\frac{10}{2}=5, x_3=\frac{f(3)}{m_3}=\frac{15}{1}=15, x_4=\frac{f(4)}{m_4}=\frac{30}{1}=30\right\}.
		\end{align*}
		Now let us construct $\mathfrak{D}$ in Construction~\ref{cons-LHSDP}. When ${\bf a}=(1,1)$, from \eqref{eq-family} we have 
		\begin{align*}
			\mathcal{D}_{1,1}=&\{(\alpha_1 a_1x_1+\alpha_2 a_2x_2)+\alpha_3 x_3+\alpha_4 x_4\ |\ \alpha_1,\alpha_2,\alpha_3,\alpha_4\in\{-1,1\}\}\\
			=&\{\alpha_1 \cdot 1 + \alpha_2 \cdot 5 + \alpha_3 \cdot 15 +\alpha_4 \cdot 30 \ |\ \alpha_1,\alpha_2,\alpha_3,\alpha_4\in\{-1,1\} \}\\
			=&\{51, -9, 21, -39, 41, -19, 11, -49, 49, -11, 19, -41, 39, -21, 9, -51\} \\ 
			=&\{\pm 9, \pm 11, \pm 19, \pm 21, \pm 39, \pm 41, \pm 49, \pm 51\}.
		\end{align*}For instance, the integer $41$ in  $\mathcal{D}_{(1,1)}$ can be obtained by $1-5+15+30=41$. Similarly, we can obtain the following blocks. 
		\begin{align*}
			\mathfrak{D}=\{
			&\mathcal{D}_{1,1}=\{\pm 9, \pm 11, \pm 19, \pm 21, \pm 39, \pm 41, \pm 49, \pm 51\},\ \ \ \ \mathcal{D}_{2,1}=\{\pm 8, \pm 12, \pm 18, \pm 22, \pm 38, \pm 42, \pm 48, \pm 52\},\\
			&\mathcal{D}_{1,2}=\{\pm 4, \pm 6, \pm 24, \pm 26, \pm 34, \pm 36, \pm 54, \pm 56\},\ \ \ \ \ \mathcal{D}_{2,2}=\{\pm 3, \pm 7, \pm 23, \pm 27, \pm 33, \pm 36, \pm 53, \pm 57\}\}.
		\end{align*}
		
		Clearly, $\mathfrak{D}$ has $m_1m_2m_3m_4=4$ blocks. The minimum and maximum integers of $\mathfrak{D}$ are $-57$ and $57$, respectively. To ensure that these two sets $[-57:0]$ and $[1:57]$ do not have overlap in $\mathbb{Z}_v$, we set $v \geq 2\times 57+1=115$. In this case we have that the intersection of any blocks is empty, and each block has $16$ different integers. 
		
		Let us take $51$ as an example, that is when $i=1$ and $j\in [2:16]$ in Definition \ref{def-LHSDP}, from \eqref{eq-HS} we have 
		\begin{equation}\label{half-sum-set}
			\begin{split}
				\mathcal{B}_{1,1} =&\left\{\frac{51+(-9)}{2}=1 \cdot 1+ 1\cdot 5 +1\cdot 15 +0 \cdot 30=21, \ \ \frac{51+21}{2}=1 \cdot 1+ 1\cdot 5 +0\cdot 15 +1 \cdot 30=36,\right.\\
				&\ \ \frac{51+(-39)}{2}=1 \cdot 1+ 1\cdot 5 +0\cdot 15 +0 \cdot 30=6, \ \ \frac{51+41}{2}=1 \cdot 1+ 0\cdot 5 +1\cdot 15 +1 \cdot 30=46,\\
				&\ \ \frac{51+(-19)}{2}=1 \cdot 1+ 0\cdot 5 +1\cdot 15 +0 \cdot 30=16, \ \ \frac{51+11}{2}=1 \cdot 1+ 0\cdot 5 +0\cdot 15 +1 \cdot 30=31,\\
				&\ \ \frac{51+(-49)}{2}=1 \cdot 1+ 0\cdot 5 +0\cdot 15 +0 \cdot 30=1, \ \ \frac{51+49}{2}=0 \cdot 1+ 1\cdot 5 +1\cdot 15 +1 \cdot 30=50,\\
				&\ \ \frac{51+(-11)}{2}=0 \cdot 1+ 1\cdot 5 +1\cdot 15 +0 \cdot 30=20, \ \ \frac{51+19}{2}=0 \cdot 1+ 1\cdot 5 +0\cdot 15 +1 \cdot 30=35,\\
				&\ \ \frac{51+(-41)}{2}=0 \cdot 1+ 1\cdot 5 +0\cdot 15 +0 \cdot 30=5, \ \ \frac{51+39}{2}=0 \cdot 1+ 0\cdot 5 +1\cdot 15 +1 \cdot 30=45,\\
				&\ \ \frac{51+(-21)}{2}=0 \cdot 1+ 0\cdot 5 +1\cdot 15 +0 \cdot 30=15, \ \ \frac{51+9}{2}=0 \cdot 1+ 0\cdot 5 +0\cdot 15 +1 \cdot 30=30,\\
				&\ \ \left.\frac{51+(-51)}{2}=0 \cdot 1+ 0\cdot 5 +0\cdot 15 +0 \cdot 30=0\right\}=\{0,1,5,6,15,16,20,21,30,31,35,36,45,46,50\}.
			\end{split}
		\end{equation} 
		By counting the intersection between $\mathcal{B}_{1,1}$ and each block of $\mathfrak{D}$, we have that $$\sum_{{\bf a} \in \mathcal{A}}|\mathcal{D}_{\bf a} \cap \mathcal{B}_{1,1}| = |\{6,21,36\}| = 3 < 4=L.$$ Similarly by checking the other half-sum set of generated by  each block of $\mathfrak{D}$, we have that $(\mathbb{Z}_{115},\mathfrak{D})$ is an $4$-$(115,16,4)$ HSDP.
		\begin{remark}\rm
			From \eqref{half-sum-set}, we can observe that the coefficient vectors corresponding to the half-sums $6$, $21$, and $36$ have all nonzero entries in the first $n=2$ coordinates, while at least one of the last $r=2$ coordinates is zero. This behavior is ensured by the following property.
			\begin{itemize}
				\item Tail--Zero Property: If a half-sum is contained in a block, then its coefficient vector must contain at least one zero among the last $r$ coordinates (this property also implies that the first $n$ coefficients are all nonzero; see Appendix~\ref{proof of Optimal} for details).
			\end{itemize}
			The Tail--Zero property imposes structural constraints on the coefficient representations of half-sums, which allows the intersection between the half-sum set and the blocks to be as large as possible, while still satisfying the $L$-half-sum condition.
		\end{remark}

		By Theorem \ref{MAPDA via LHSDP} we have a $(4,115,115,51,460)$ MAPDA which generates a $F$-division $(L=4,K=115,M,N)$ MISO coded caching scheme with memory ratio $M/N=51/115$, the sum-DoF $g = 16$ and subpacketization $F=K=115$. Now let us consider the existing schemes with $K=115$ and $M/N=51/115$.
		\begin{itemize}
			\item The YWCC $1$ scheme: In \cite{YWCC} we have a $(L = 4, K=115,M,N)$ YWCC $1$ scheme with the memory ratio $M/N = 51/115$, the subpacketization $F_{\text{YWCC}} = 55\binom{115}{51} \approx8.13 \times 10^{34} \gg 115=F$, the sum-DoF $g_{\text{YWCC}} = 55 > 16 = g$. So our scheme achieves a little lower sum-DoF, but significantly reduces the subpacketization compared to the YWCC $1$ scheme;
			\item  The CTWWL scheme: In \cite{CTWWL} we have a $(L=4,K=115,M,N)$ CTWWL scheme with the memory ratio $M/N=51/115$, the subpacketization $F_{\text{CTWWL}} = 960 > 115 = F$, the sum-DoF $g_{\text{CTWWL}} = 8 < 16 = g$. So our scheme has lower subpacketization but larger sum-DoF while maintaining the same memory ratio.
		\end{itemize}
		\hfill $\square$  
	\end{example}
	
	Now, let us consider the value of $v$ to ensure that the resulting pair in Construction~\ref{cons-LHSDP} is an HSDP. For any positive integers $n$ and $m_1$, $m_2$, $\ldots$, $m_{n}$, suppose the vector 
	\begin{align*}
		\mathbf{a}=(a_1,a_2,\ldots,a_{n})=(m_1,m_2,\ldots,m_{n}) \in \mathcal{A}.
	\end{align*}In Construction~\ref{cons-LHSDP}, we have the following block  
	\begin{align*}
		\mathcal{D}_{\mathbf{a}}=\{\alpha_1 f(1)+\alpha_2 f(2)+\cdots+\alpha_{n+r} f(n+r)\  |\  \alpha_i \in \{-1,1\}, i \in [n+r]\}, 
	\end{align*}which contains the integers   $-f(1)-f(2)-\dots-f(n+r)$ and $f(1)+f(2)+\ldots+f(n+r)$. To ensure that  $[-f(1)-f(2)-\dots-f(n+r):0]$ and $[1:f(1)+f(2)+\ldots+f(n+r)]$ can not overlap in $\mathbb{Z}_{v}$, the value of $v$ must satisfy $$v\geq 2(f(1)+f(2)+\ldots+f(n+r))+1.$$
	For the ease of further notations, we define that
	\begin{align*}
		\phi(m_1,m_2,\ldots,m_n):=&\sum_{i=1}^{n+r}f(i) = \sum_{i=1}^{n}f(i) + \sum_{i=n+1}^{n+r}f(i).
	\end{align*}
	Since 
	\begin{equation}
		\begin{aligned}
			\label{sum_f_n+1}
			\sum_{i=n+1}^{n+r}f(i)=&(1+2+4+\dots+2^{r-1}+2^r-L)f(n+1)=(2^{r+1}-L-1)f(n+1)\\
			=&(2^{r+1}-L-1)(1+m_{n})\frac{f(n)}{m_{n}}=(2^{r+1}-L-1)(1+m_{n})\left(1+2\sum_{i=1}^{n-1}f(i)\right),
		\end{aligned}
	\end{equation}
	leveraging the recursively defined property by Construction~\ref{cons-LHSDP}, we have
	\begin{equation}
		\begin{aligned}
			\label{sum_f_n}
			\sum_{i=1}^{n}f(i)=&\sum_{i=1}^{n-1}f(i)+m_{n}\left(\sum_{j=1}^{n-1}2f(j)+1\right)=(1+2m_{n})\sum_{i=1}^{n-1}f(i)+m_{n}\\
			=&(1+2m_{n})(1+2m_{n-1})\sum_{i=1}^{n-2}f(i)+m_{n-1}(1+2m_{n})+m_{n}\\
			=&\prod_{i=n-2}^{n}(1+2m_i)\sum_{i=1}^{n-3}f(i)+m_{n-2}\prod_{i=n-1}^{n-r}(1+2m_i)+m_{n-1}(1+2m_{n})+m_{n}\\
			=&\sum_{i=1}^{n-1}\left(m_i\prod_{j=i+1}^{n}(1+2m_{i})\right)+ m_{n}.
		\end{aligned}
	\end{equation}
	Combining (\ref{sum_f_n+1}) and (\ref{sum_f_n}), we can obtain
	\begin{equation}
		\begin{aligned}
			\label{eq_sum_f}
			\phi(m_1,m_2,\ldots,m_{n}):=&\sum_{i=1}^{n+r}f(i) = \sum_{i=1}^{n}f(i) + \sum_{i=n+1}^{n+r}f(i)\\
			=&\sum_{i=1}^{n-1}\left(m_i\prod_{j=i+1}^{n}(1+2m_{i})\right)+ m_{n} + (2^{r+1}-L-1)(1+m_{n})\\
			&+ (2^{r+2}-2L-2)(1+m_{n})\left(\sum_{i=1}^{n-2}\left(m_i\prod_{j=i+1}^{n-1}(1+2m_{i})\right)+ m_{n-1}\right)\\
			=&\left(2(2^{r+1}-L)(1+m_{n}) -1\right)\left(\sum_{i=1}^{n-2}\left(m_i\prod_{j=i+1}^{n-1}(1+2m_{i})\right) + m_{n-1}\right)\\
			&+(2^{r+1}-L)(1+m_{n})-1.
		\end{aligned}
	\end{equation}
	From \eqref{eq_sum_f} and by Theorem \ref{MAPDA via LHSDP}, the following main result can be obtained. In addition, its detailed proof is included in Appendix~\ref{proof of Optimal}.
	\begin{theorem}\rm
		\label{th-main-theorem-MAPDA}
		For any positive integers $L$, $r$ and $m_1$, $m_2$, $\ldots$, $m_{n}$ satisfying $L \leq 2^r$, the pair $(\mathbb{Z}_{v},\mathfrak{D})$ generated in Construction~\ref{cons-LHSDP} is an $L$-$(v,2^{n+r},\prod_{i=1}^{n}m_i)$ HSDP under the constraint of $v\geq 2\phi(m_1,m_2,\ldots,m_{n})+1$. Then we can obtain an $(L,v,v,Z=v-2^{n+r}\prod_{i=1}^{n}m_i,S=v\prod_{i=1}^{n}m_i)$ MAPDA which generates an $(L,v,M,N)$ MISO coded caching scheme with memory ratio $M/N=1-(2^{n+r}\prod_{i=1}^{n}m_i)/v$ and sum-DoF $g=2^{n+r}$.
		\hfill $\square$  
	\end{theorem}
	
	By Theorem~\ref{def-LHSDP}, for given parameters $v$, $r$, and $n$ (recall that $K = v$ and $g = 2^{n+r}$), 
	we focus on selecting $m_1, \ldots, m_n$ so as to minimize the memory ratio. 
	In other words, with the coded caching gain fixed, our objective is to find the coded caching scheme achieving the minimal memory ratio. By including the constraint in the optimization problem for choosing $m_1, m_2, \ldots, m_n$, we have
	\begin{equation}\label{eq-optimization}
		\begin{aligned}
			\text{Problem 1.}  \ \ \  & \textbf{Maximize } \text{function } f=\prod_{i=1}^{n}m_i \\
			&\textbf{Constrains: } m_1,\ldots,m_{n} \in \mathbb{Z}^{+},\\
			&\quad\quad\quad \quad \quad  \left(2(2^{r+1}-L)(1+m_{n}) -1\right)\left(\sum_{i=1}^{n-2}\left(m_i\prod_{j=i+1}^{n-1}(1+2m_{i})\right) + m_{n-1}\right) \\
			&\quad\quad\quad \quad \quad+ (2^{r+1}-L)(1+m_{n})-1 \leq \frac{v-1}{2}.
		\end{aligned}
	\end{equation}
	Problem~1 is an integer programming problem known to be NP-hard. To simplify the problem, we first relax the integer constraint on $m_1, m_2, \ldots, m_n$ from $m_i \in \mathbb{Z}^+$ to $m_i \in \mathbb{R}^+$, for $i \in [n]$. This relaxation transforms the problem into a convex optimization problem, which is then solved using the Lagrange Multiplier Method.
	
	The following theorem introduces the sub-optimal solution by the above optimization strategy, whose proof is given in Appendix~\ref{sec:proof of th_sub_operation}.
	
	\begin{theorem}\rm
		\label{th_sub_operation}
		A sub-optimal solution (with closed-form) to Problem~1 is 
		\begin{align}
			\label{eq:m1_m_{n}}
			q=m_1=m_2=\cdots=m_{n-1},\ \  m_{n} = \frac{(2^{r+2}-2L-1)q}{2^{r+1}-L},
		\end{align}
		where $q$ is a positive integer. Under the selection in~\eqref{eq:m1_m_{n}}, we can obtain the $L$-$((2^{r+2}-2L-1)(1+2q)^n, 2^{n+r}, (2^{r+2}-2L-1)q^n/(2^{r+1}-L))$ HSDP and the $(L,K,F,Z,S)$ MAPDA in Theorem \ref{th-main-theorem-MAPDA} where    
		\begin{align*}
			K&=F=v=(2^{r+2}-2L-1)(1+2q)^n,\ \   S=\frac{(2^{r+2}-2L-1)^{2}q^{n}(1+2q)^{n}}{2^{r+1}-L}, \\
			Z&=(2^{r+2}-2L-1)(1+2q)^n\left(1-\frac{2^r}{2^{r+1}-L}\left(\frac{2q}{1+2q}\right)^n\right), 
		\end{align*} which can realize a coded caching scheme with the memory ratio $M/N=1-2^r(2q/(1+2q))^n/(2^{r+1}-L)$ and the sum-DoF $g=2^{n+r}$.
		\hfill $\square$ 
	\end{theorem}
	When $L=2^r$, we have $v=(2^{r+1}-1)(1+2q)^n$ in Theorem \ref{th_sub_operation}. Then the following result can be obtained.
	\begin{corollary}[Sub-optimal solution of Problem 1 with $L=2^r$]\rm
		\label{corollary-L=2^r}
		For any positive integers $q$, $n$ and $r$, there exists a $2^r$-$((2^{r+1}-1)(1+2q)^n,2^{n+r},(2^{r+1}-1)q^n/2^r)$ HSDP. Then we have a $(L = 2^r, K=(2^{r+1}-1)(1+2q)^n, F=(2^{r+1}-1)(1+2q)^n,Z=(2^{r+1}-1)\left((1+2q)^n-(2q)^n\right),S=(2^{r+1}-1)^{2}q^{n}(1+2q)^{n}/2^{r})$ MAPDA which realizes an $(L=2^r,K=(2^{r+1}-1)(1+2q)^n,M,N)$ MISO coded caching scheme with the subpacketization $F=(2^{r+1}-1)(1+2q)^n$, the memory ratio $M/N=1-(2q/(1+2q))^n$ and sum-DoF $g=2^{n+r}$. 
	\end{corollary}
	
	\section{PERFORMANCE ANALYSIS}
	\label{sec-perf-ana}
	In this section, we will present theoretical and numerical comparisons with the existing schemes \cite{YWCC, CTWWL,WCC,NPR}, which are  summarized in Table~\ref{table_existing_schemes}, respectively to show the performance of our new scheme in Theorem \ref{th_sub_operation}. 
	
	\subsection{Theoretical comparisons }
	Since the expressions for user number and memory ratio in Theorem \ref{th_sub_operation} are complex, we focus on the special case $L=2^r$, as presented in Corollary \ref{corollary-L=2^r}, for comparison with existing schemes with linear subpacketization \cite{WCC,CTWWL} and exponential subpacketization \cite{YWCC, NPR} respectively.
	
	\subsubsection{Comparison with the WCC scheme in \cite{WCC}}
	In Table \ref{table_existing_schemes}, when $L = 2^r$, $K=(2^{r+1}-1)(1+2q)^n$ we have the $(L=2^r, K=(2^{r+1}-1)(1+2q)^n,M,N)$ WCC scheme with the memory ratio $M/N=1-(2q/(1+2q))^n$, subpacketization $F_{\text{WCC}}$ in Table \ref{table_existing_schemes}, and the sum-DoF $g_{\text{WCC}} = 2^{r+1}$. Then the following ratios  
	\begin{align*}
		\frac{F_{\text{WCC}}}{F}\geq  \frac{(2^{k+1}-1)(1+2q)^n}{(2^{k+1}-1)(1+2q)^n}=1 
		\ \ \text{and} \ \ \frac{g_{\text{WCC}}}{g}=\frac{2^{r+1}}{2^{n+r}}=\frac{1}{2^{n-1}}.
	\end{align*}can be obtained. 
	Compared to the WCC scheme, our scheme either obtaining the same subpacketization or the reduction amount of our subpacketization is larger than 2 times. The multiplicative increase in our sum-DoF is at least $2^{n-1}$.
	
	\subsubsection{Comparison with the CTWWL scheme in \cite{CTWWL}}
	In Table \ref{table_existing_schemes}, when $L = 2^r$, $K=(2^{r+1}-1)(1+2q)^n$ we have the $(L=2^r, K=(2^{r+1}-1)(1+2q)^n,M,N)$ CTWWL scheme with the memory ratio $M/N=1-(2q/(1+2q))^n$. Since $t+L \geq K-t+L$, so we have 
	\begin{align}
		\label{eq_tL_KtL}
		(2^{r+1}-1)((1+2q)^n-(2q)^n)+2^r = x((2^{r+1}-1)(2q)^n+2^r) + y, x \geq 1, 2^r \leq y \leq (2^{r+1}-1)(2q)^n+2^r
	\end{align}
	Then the following results can be obtained.
	\begin{itemize}
		\item If $y < 2L=2^{r+1}$, we have $g_{\text{CTWWL}} = 2^{r+1}x+y$. Since $2^r \leq y$, so from \eqref{eq_tL_KtL}, we have $x \leq \frac{(2^{r+1}-1)((1+2q)^n-(2q)^n)}{((2^{r+1}-1)(2q)^n+2^r)}$, then the following ratios can be obtained.
		\begin{align*}
			\frac{F_{\text{CTWWL}}}{F}
			=&\frac{g_{\text{CTWWL}}F}{F} = g_{\text{CTWWL}} = 2^{r+1}x+y \geq 2^{r+1},
			\\
			\frac{g_{\text{CTWWL}}}{g}
			=&\frac{2^{r+1}x+y}{2^{n+r}} < \frac{2^{r+1}x+2^{r+1}}{2^{n+r}} = \frac{x+1}{2^{n-1}} \leq \frac{(2^{r+1}-1)(1+2q)^n + 2^r}{2^{n-1}((2^{r+1}-1)(2q)^n+2^r)} \\
			< & \frac{(2^{r+1}-1)(1+2q)^n + 2^r}{2^{n-1}(2^{r+1}-1)(2q)^n}
			< \frac{1}{2^{n-1}}\left((1+\frac{1}{2q})^n+\frac{1}{(2-\frac{1}{2^{r}})(2q)^n}\right)\\
			< & \frac{1}{2^{n-1}}\left((1+\frac{1}{2q})^n+\frac{1}{(2q)^n}\right)
			< \frac{1}{2^{n-1}}\left((1+\frac{1}{2})^n + \frac{1}{2^n}\right) \\
			=& \frac{3^n+1}{2^{2n-1}}.
		\end{align*}
		
		\item If $y \geq 2L=2^{r+1}$, we have $g_{\text{CTWWL}} = 2^{r+1}(x+1)$, Since $2^{r+1} \leq y$, so from \eqref{eq_tL_KtL}, we have $x \leq \frac{(2^{r+1}-1)((1+2q)^n-(2q)^n)-2^r}{((2^{r+1}-1)(2q)^n+2^r)}$, then the following ratios can be obtained.
		\begin{align*}
			\frac{F_{\text{CTWWL}}}{F}
			=&\frac{g_{\text{CTWWL}}F}{F} = g_{\text{CTWWL}} = 2^{r+1}(x+1) \geq 2^{r+2},\\
			\frac{g_{\text{CTWWL}}}{g}
			=&\frac{2^{r+1}(x+1)}{2^{n+r}} = \frac{x+1}{2^{n-1}} \leq \frac{(2^{r+1}-1)(1+2q)^n}{2^{n-1}((2^{r+1}-1)(2q)^n+2^r)} \\
			< & \frac{(2^{r+1}-1)(1+2q)^n}{2^{n-1}(2^{r+1}-1)(2q)^n}
			< \frac{1}{2^{n-1}}(1+\frac{1}{2q})^n 
			\leq \frac{1}{2^{n-1}}(1+\frac{1}{2})^n \\
			=& \frac{3^n}{2^{2n-1}}.
		\end{align*}
	\end{itemize}
	As a result, compared to the CTWWL scheme in \cite{CTWWL}, the reduction amount of our subpacketization is larger than $2^{r+2}$ times. The multiplicative increase in our sum-DoF is at least $\frac{2^{2n-1}}{3^n+1}$. When $n \geq 3$, the formula $\frac{2^{2n-1}}{3^n+1} >XQ 1$ always holds.
	
	\subsubsection{Comparison with the YWCC 1 scheme in \cite{YWCC}}
	In Table \ref{table_existing_schemes}, when $L = 2^r$, $K=(2^{r+1}-1)(1+2q)^n$ we have the $(L=2^r, K=(2^{r+1}-1)(1+2q)^n,M,N)$ YWCC scheme with the memory ratio $M/N=1-(2q/(1+2q))^n$, the sum-DoF $g_{\text{YWCC}} =KM/N+L = \left((2^{r+1}-1)\left((1+2q)^n - (2q)^n\right)+2^r\right)$. Let us consider the ratio of the sum-DoF between $g$ in Corollary \ref{corollary-L=2^r} and the YWCC scheme as follows.
	\begin{align*}
		\frac{g_{\text{YWCC}}}{g}
		=&\frac{(2^{r+1}-1)\left((1+2q)^n-(2q)^n\right)+2^r}{2^{n+r}}=\frac{2^{r+1}(1-\frac{1}{2^{r+1}})\cdot (2q)^n\left((1+\frac{1}{2q})^n-1\right)+2^r}{2^{n+r}}\\
		=&2q^n\left(1-\frac{1}{2^{r+1}}\right)\left(\left(1+\frac{1}{2q}\right)^n-1\right)+\frac{1}{2^{n}} \\
		<&2q^n\left(\left(1+\frac{1}{2q}\right)^n-1\right)+\frac{1}{2^{n}}.
	\end{align*}
	For any given integers $q$ and $n$ we have that the ratio of sum-DoFs $\frac{g_{\text{YWCC}}}{g}$ is less than a fixed real number $2q^n((1+\frac{1}{2q})^n-1)+\frac{1}{2^{n}}$. Now, let us consider the ratio of the subpacketizations between in Corollary \ref{corollary-L=2^r} and the YWCC scheme. By Table \ref{table_existing_schemes}, the subpacketization $F_{\text{YWCC}}$ changes when the value of the parameter $m$ changes. So, in order to clearly demonstrate the reduction in the level of subpacketisation achieved by our scheme compared to the YWCC scheme, we will only consider case $m = 1$. Specifically, by Table \ref{table_existing_schemes} when  $m=1$ we have the subpacketization of the YWCC scheme as follows.
	\begin{align*}
		F_{\text{YWCC}}=&(t+L){K\choose t}=\left((2^{r+1}-1)\left((1+2q)^n - (2q)^n\right)+2^r\right)\binom{(2^{r+1}-1)(1+2q)^n}{(2^{r+1}-1)\left((1+2q)^n - (2q)^n\right)} \\
		>&\left((2^{r+1}-1)\left((1+2q)^n - (2q)^n\right)+2^r\right)\left(1+\frac{1}{2q}\right)^{n(2^{r+1}-1)(2q)^n}.
	\end{align*}
	Then the following subpacketization ratio can be obtained.
	\begin{align*}
		\frac{F_{\text{YWCC}}}{F}
		=&\frac{\left((2^{r+1}-1)\left((1+2q)^n - (2q)^n\right)+2^r\right)\binom{(2^{r+1}-1)(1+2q)^n}{(2^{r+1}-1)(2q)^n}}{(2^{r+1}-1)(1+2q)^n}\\
		=&\left(1-\frac{(2q)^n}{(1+2q)^n} + \frac{2^r}{(2^{r+1}-1)(1+2q)^n}\right)\binom{(2^{r+1}-1)(1+2q)^n}{(2^{r+1}-1)(2q)^n}\\
		>&\left(1-\frac{1}{(1+\frac{1}{2q})^n}\right)\left(1+\frac{1}{2q}\right)^{n(2^{r+1}-1)(2q)^n}\\
		=&\left(1-\frac{1}{(1+\frac{1}{2q})^n}\right)\left(1+\frac{1}{2q}\right)^{\frac{nK}{(1+\frac{1}{2q})^n}}.
	\end{align*}
	We can see that the sum-DoF of our scheme in Corollary \ref{corollary-L=2^r} decreases at most $2q^n((1+\frac{1}{2q})^n-1)+\frac{1}{2^{n}}$ times while the subpacketization of our scheme in  Corollary \ref{corollary-L=2^r} decreases atleast $(1-\frac{1}{(1+\frac{1}{2q})^n})(1+\frac{1}{2q})^{\frac{nK}{(1+\frac{1}{2q})^n}}$ times.
	
	\subsubsection{Comparison with the NPR scheme in \cite{NPR}}
	In Table \ref{table_existing_schemes}, when $L = 2^r$, $K=(2^{r+1}-1)(1+2q)^n$ we have the $(L=2^r, K=(2^{r+1}-1)(1+2q)^n,M,N)$ NPR scheme with the parameter $\beta=\gcd(K,t,L)=1$, the memory ratio $M/N=1-(2q/(1+2q))^n$, the sum-DoF $g = t+L = \left((2^{r+1}-1)\left((1+2q)^n - (2q)^n\right)+2^r\right)$ and the subpacketization 
	\begin{align*}
		F_{\text{NPR}}=&\frac{t+L}{\beta}\binom{\frac{K}{\beta}}{\frac{t+L}{\beta}}
		=\left((2^{r+1}-1)\left((1+2q)^n - (2q)^n\right)+2^r\right)\binom{(2^{r+1}-1)(1+2q)^n}{(2^{r+1}-1)((1+2q)^n - (2q)^n)+2^r}\\
		=&\left((2^{r+1}-1)\left((1+2q)^n - (2q)^n\right)+2^r\right)\binom{(2^{r+1}-1)(1+2q)^n}{(2^{r+1}-1)(2q)^n-2^r}\\
		>&\left((2^{r+1}-1)\left((1+2q)^n - (2q)^n\right)+2^r\right)\left(\frac{(2^{r+1}-1)(1+2q)^n}{(2^{r+1}-1)(2q)^n-2^r}\right)^{(2^{r+1}-1)(2q)^n-2^r}\\
		>&\left((2^{r+1}-1)\left((1+2q)^n - (2q)^n\right)+2^r\right)\left(1+\frac{1}{2q}\right)^{n((2^{r+1}-1)(2q)^n-2^r)}.
	\end{align*}
	Then the following ratios can be obtained.
	\begin{align*}
		\frac{F_{\text{NPR}}}{F}
		=&\frac{((2^{r+1}-1)\left((1+2q)^n - (2q)^n\right)+2^r)\binom{(2^{r+1}-1)(1+2q)^n}{(2^{r+1}-1)((1+2q)^n - (2q)^n)+2^r}}{(2^{r+1}-1)(1+2q)^n} \\
		>& \frac{\left((2^{r+1}-1)\left((1+2q)^n - (2q)^n\right)+2^r\right)}{(2^{r+1}-1)(1+2q)^n}\left(1+\frac{1}{2q}\right)^{n((2^{r+1}-1)(2q)^n-2^r)} \\
		>&\left(1-\frac{1}{(1+\frac{1}{2q})^n}\right)\left(1+\frac{1}{2q}\right)^{n((2^{r+1}-1)(2q)^n-2^r)}\\
		=&\left(1-\frac{1}{(1+\frac{1}{2q})^n}\right)\left(1+\frac{1}{2q}\right)^{\frac{nK}{(1+\frac{1}{2q})^n} - n2^r}.
	\end{align*}
	Similar to the YWCC scheme, the sum-DoF comparison between the scheme in Corollary \ref{corollary-L=2^r} and the NPR scheme is as follows.
	\begin{align*}
		\frac{g_{\text{NPR}}}{g} < 2q^n\left(\left(1+\frac{1}{2q}\right)^n-1\right)+\frac{1}{2^{n}}.
	\end{align*}
	Compared to the NPR scheme, the reduction amount of our subpacketization is about $(1-\frac{1}{(1+\frac{1}{2q})^n})(1+\frac{1}{2q})^{\frac{nK}{(1+\frac{1}{2q})^n} - n2^r}$ times while the sum-DoF of NPR scheme increases $2q^n((1+\frac{1}{2q})^n-1)+\frac{1}{2^{n}}$ times. So our scheme has the significant advantages on subpacketization compared to the NPR scheme.

	\subsection{Numerical comparisons }\label{subsec-p}
	In this subsection, we will present numerical comparisons between our scheme in Theorem \ref{th_sub_operation} and the existing schemes\cite{WCC,CTWWL,YWCC, NPR}. 
	
	\subsubsection{Comparison with the linear subpacketization schemes in \cite{CTWWL, WCC}}
\begin{figure}[htbp!]
		\centering
		\begin{minipage}[t]{0.495\textwidth}
			\centering
			\includegraphics[height=6cm, width=\linewidth, keepaspectratio]{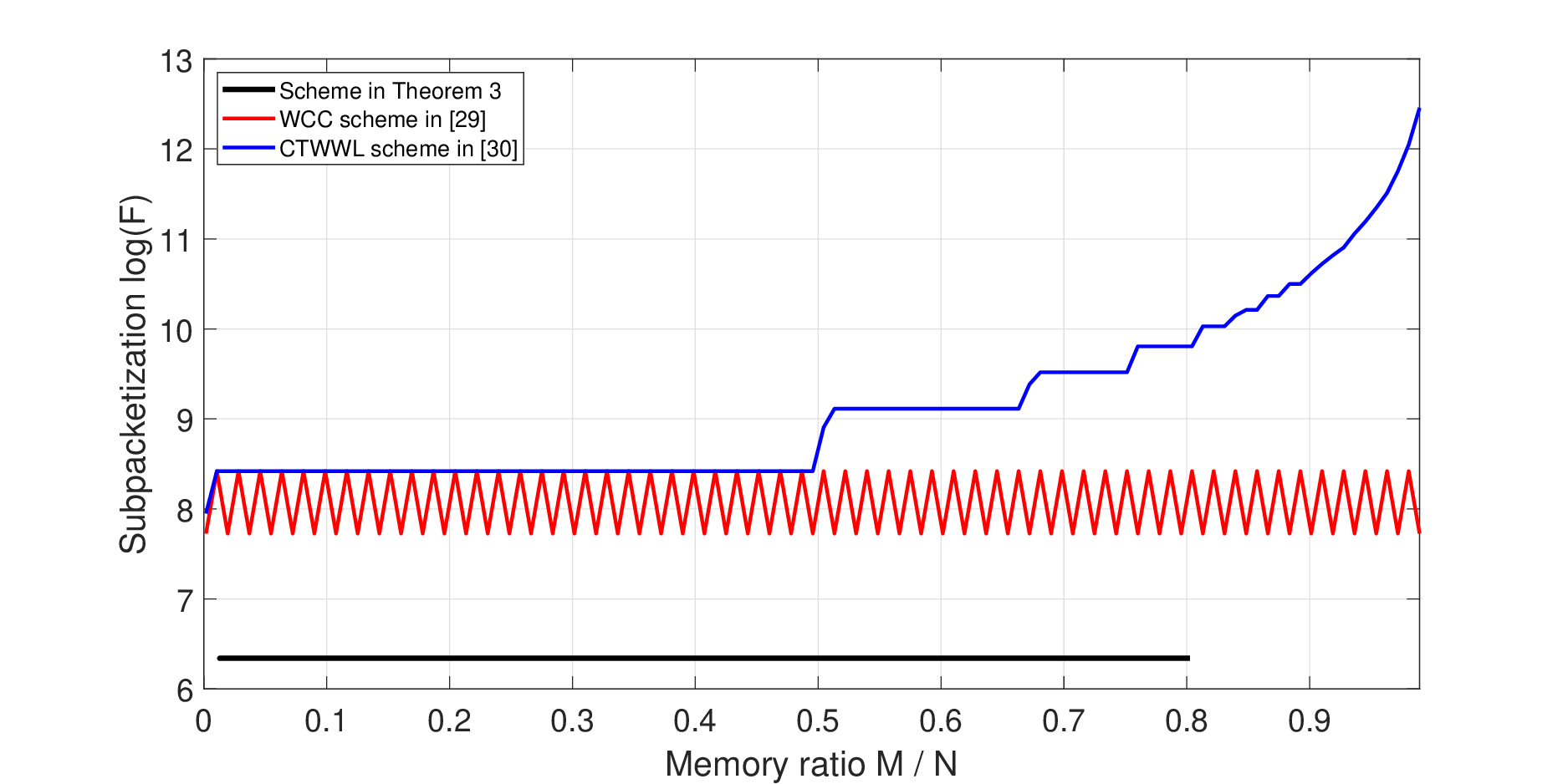}
			\caption{Memory ratio-subpacketization tradeoff for $K = 567$, $L = 4$}
			\label{fig-K=567-F}
		\end{minipage}
		\hfill
		\begin{minipage}[t]{0.495\textwidth}
			\centering
			\includegraphics[height=6cm, width=\linewidth, keepaspectratio]{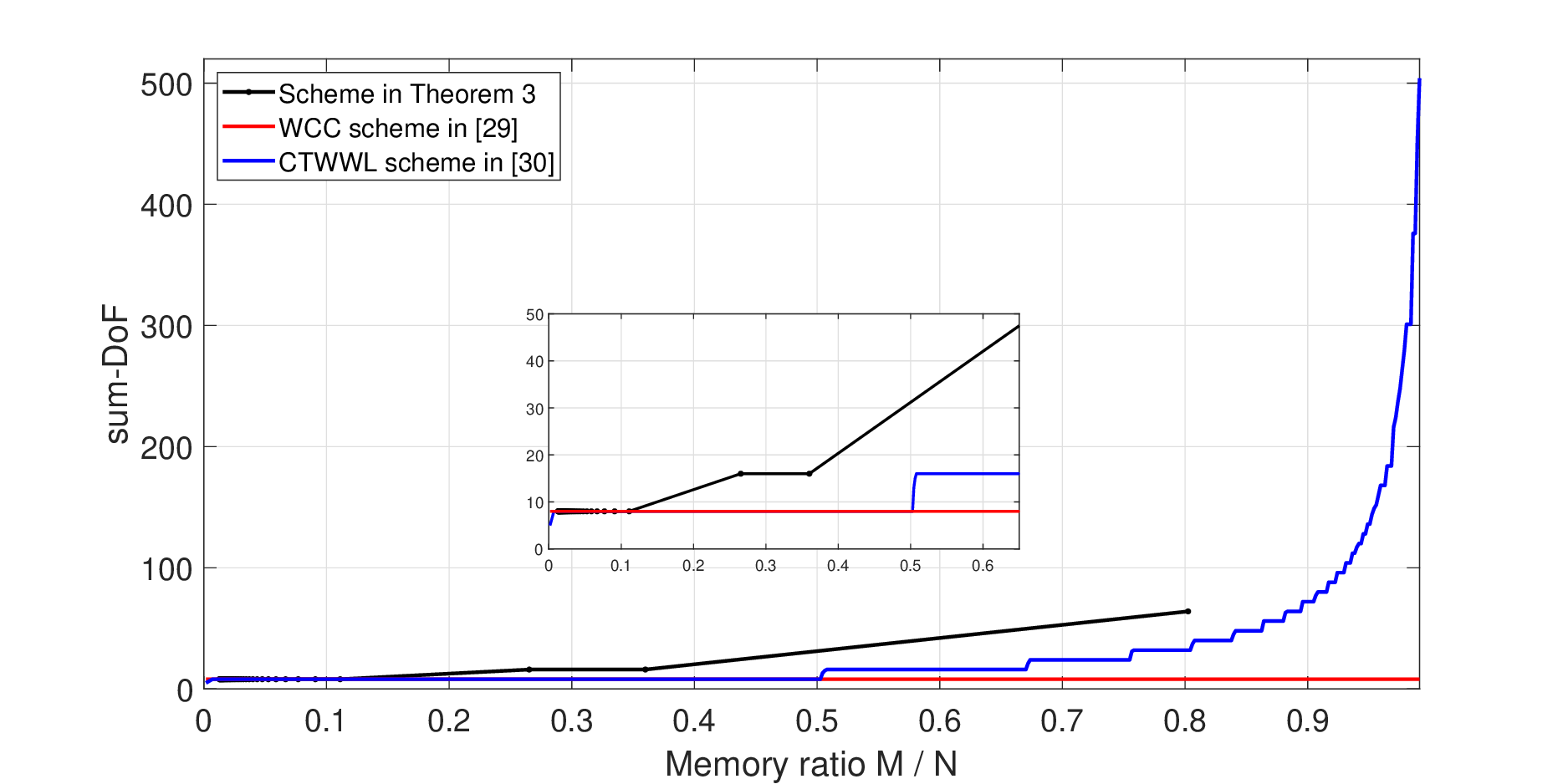}
			\caption{Memory ratio-sum-DoF tradeoff for $K = 567$, $L = 4$}
			\label{fig-K=567-DoF}
		\end{minipage}
	\end{figure}

	When $K = 567$ and $L = 4$, the performance of our scheme in Theorem \ref{th_sub_operation} (black line), the CTWWL scheme (blue line), and the WCC scheme (red line) are compared in Fig.~\ref{fig-K=567-F} and Fig.~\ref{fig-K=567-DoF}. It can be observed that our scheme achieves a lower subpacketization and higher or equal sum-DoF while maintaining the same memory ratio. The same conclusion can be drawn for $K = 567$ and $L = 4$, as shown in Fig.~\ref{fig-K=85-F} and Fig.~\ref{fig-K=85-DoF}.
	
	\subsubsection{Comparison with the exponential subpacketization schemes in \cite{YWCC, NPR}}
\begin{figure}[htbp!]
		\centering
		\begin{minipage}[t]{0.495\textwidth}
			\centering
			\includegraphics[height=6cm, width=\linewidth, keepaspectratio]{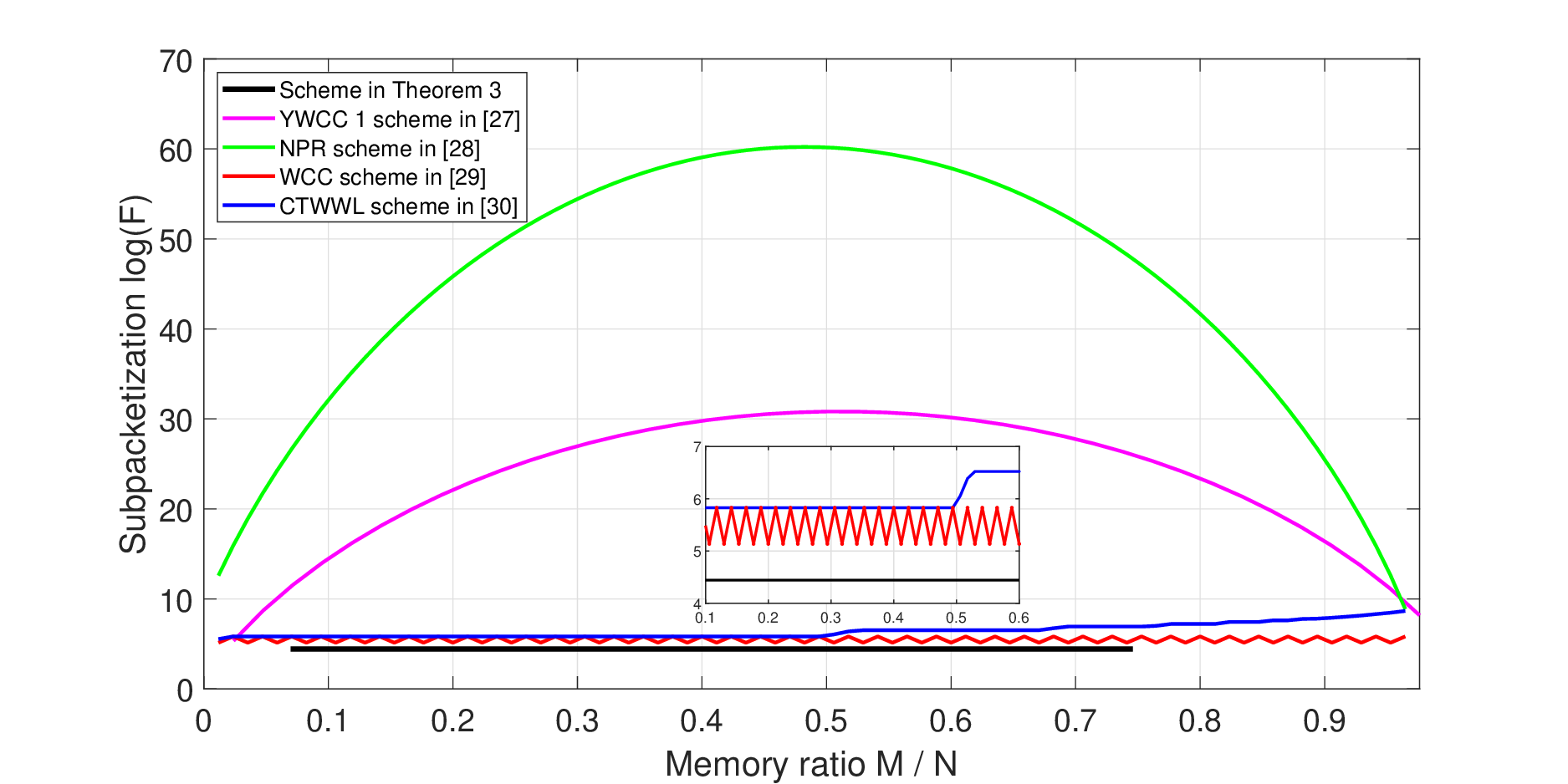}
			\caption{Memory ratio-subpacketization tradeoff for $K = 85$, $L = 2$}
			\label{fig-K=85-F}
		\end{minipage}
		\hfill
		\begin{minipage}[t]{0.495\textwidth}
			\centering
			\includegraphics[height=6cm, width=\linewidth, keepaspectratio]{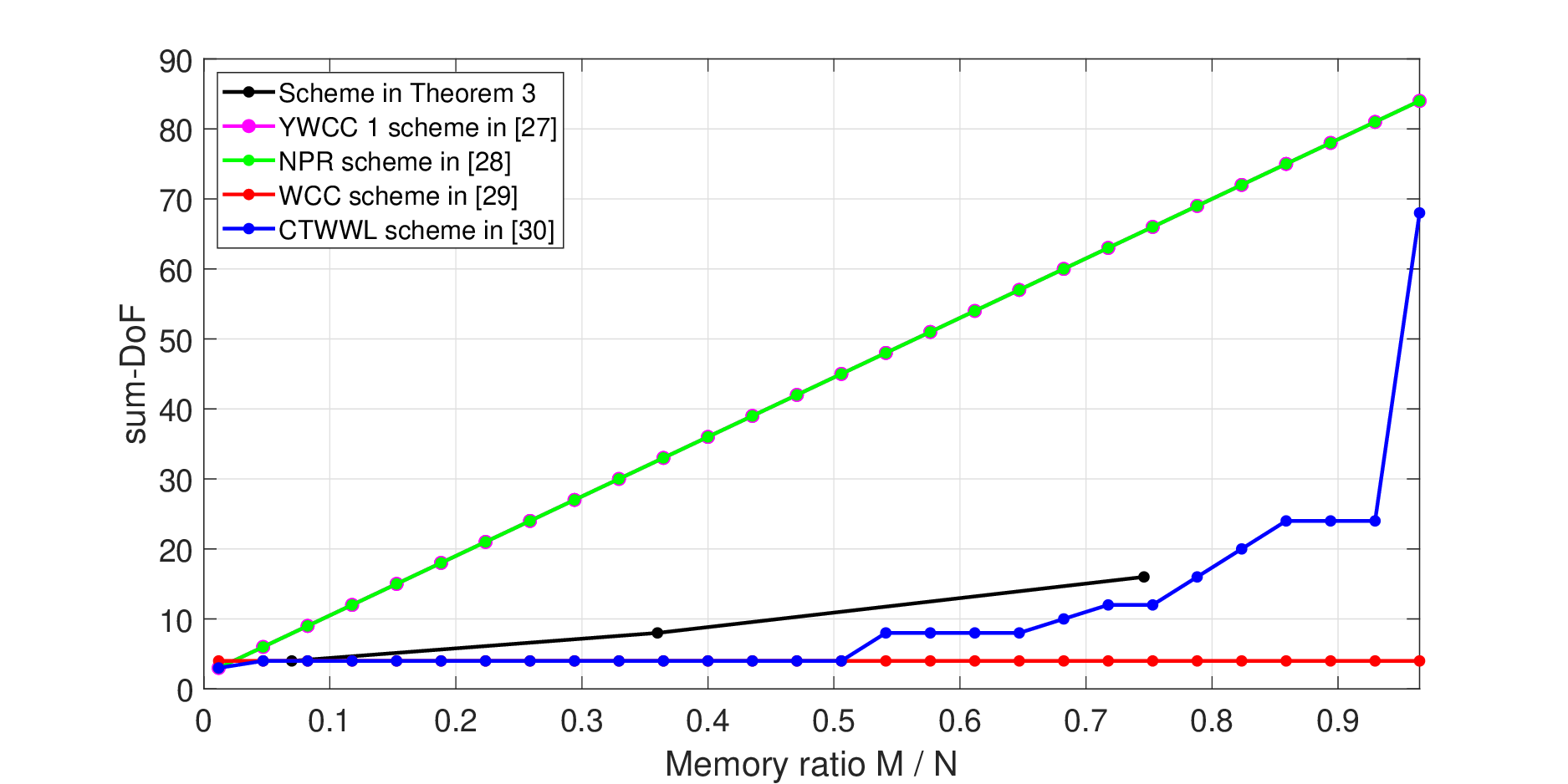}
			\caption{Memory ratio-sum-DoF tradeoff for $K = 85$, $L = 2$}
			\label{fig-K=85-DoF}
		\end{minipage}
	\end{figure}

	When $K = 85$ and $L = 2$, Fig.~\ref{fig-K=85-F} and Fig.~\ref{fig-K=85-DoF} compare our scheme in Theorem \ref{th_sub_operation} (black line), the YWCC $1$ scheme with $m=L$ (pink line), and the NPR scheme (green line). Our scheme achieves significantly lower subpacketization with only a slight reduction in sum-DoF. So our scheme achieves a favorable trade-off: a small sacrifice in sum-DoF for exponential improvements in subpacketization.
	
	\section{CONCLUSION}\label{sec-conclu}
	In this paper, we focus on MISO coded caching schemes that achieve linear subpacketization while maximizing the sum-DoF. Specifically, we introduce a novel combinatorial structure, referred to as the $L$-half-sum disjoint packing (HSDP), which enables the construction of schemes with subpacketization level $F=K$. Then by constructing $L$-HSDPs, we obtain a new class of MISO coded caching schemes with $F=K$. By theoretical and numerical analyses, demonstrate that the proposed schemes achieve higher sum-DoF and lower subpacketization than the existing schemes with linear subpacketization. Compared to some existing schemes with exponential subpacketization, our schemes achieve a significant reduction in subpacketization while only slightly sacrificing sum-DoF.
	
	This work specifically concentrated on $L$-$(v,g,b)$ HSDPs with $g = 2^{n+r}$, where $L \leq 2^r$ and $n$ is a positive integer. In future work, exploring more general constructions of HSDPs with alternative values of $g$ represents a promising direction for further research.
	
	\appendices
	\section{Proof of Theorem~\ref{MAPDA via LHSDP}}
	\label{proof of MAPDA to LHSDP}
	
	Let $(\mathbb{Z}_v,\mathfrak{D})$ be an $L$-$(v,g,b)$ HSDP, where $\mathfrak{D}=\{\mathcal{D}_1,\mathcal{D}_2,\ldots,\mathcal{D}_b\}$. According to Definition~\ref{def-MAPDA}, we first determine the number of star entries in each column. For any column $k$ and any integer $x \in \mathfrak{D}$, there exists a unique integer $f$ such that $k-f = x$. Consequently, each column contains exactly $bg$ integer entries, and thus $v - bg$ star entries. So we obtain $Z = v - bg$.
	
	We now determine the number of distinct integer pairs in $\mathbf{P}$. For any integers $i\in [b]$, $c\in\mathbb{Z}_v$ and for each integer $d\in\mathcal{D}_i$, when $v$ is odd, the following system of equations always has a unique solution:
	\begin{align*}
		\left\{
		\begin{array}{c}
			k-f=d;\\
			k+f=c. 
		\end{array}
		\right.
	\end{align*}When $d$ takes all values in $\mathcal{D}_i$, exactly $g$ distinct solutions are obtained. So the integer pair $(c,i)$ appears in $\mathbf{P}$ precisely $g$ times, and the total number of distinct integer pairs is $bv$.
	
	Finally, let us consider the Condition C$4$ of Definition \ref{def-MAPDA}. Assume that the entries $\mathbf{P}(f_1,k_1), \mathbf{P}(f_2,k_2), \dots, \mathbf{P}(f_g,k_g)$ satisfy that $\mathbf{P}(f_1,k_1)=\mathbf{P}(f_2,k_2)=\dots=\mathbf{P}(f_g,k_g) = (c,i)$. Define $\mathbf{P}^{(c,i)}$ as the subarray of $\mathbf{P}$ including the rows and columns containing $(c,i)$ and $r_s'\times r_s$ as the dimension of $\mathbf{P}^{(c,i)}$. Without loss of generality, we take $\mathbf{P}(f_1,k_1)$ as the entry point to conduct the proof. From \eqref{eq-cons-1} we have 
	\begin{align}
		\label{eq-k f}
		\begin{cases}
			d_1=k_1-f_1 \in \mathcal{D}_i, \\
			d_2=k_2-f_2 \in \mathcal{D}_i, \\
			\ \ \ \ \ \ \ \ \ \ \vdots \\
			d_g=k_g-f_g \in \mathcal{D}_i,
		\end{cases}
		\ \ \text{and} \ \ 
		\begin{cases}
			k_1+f_1=k_2+f_2, \\
			k_1+f_1=k_3+f_3, \\
			\ \ \ \ \ \ \ \ \ \ \vdots \\
			k_1+f_1=k_g+f_g.
		\end{cases}
	\end{align}
	If $f_i=f_j$ or $k_i=k_j$ for some $i \neq j$ with $i,j \in [g]$, then we must have $k_i=k_j$ or $f_i=f_j$ respectively, contradicting the hypothesis that $\mathbf{P}(f_i,k_i)$ and $\mathbf{P}(f_j,k_j)$ are distinct entries. So each integer pair in $\mathbf{P}$ appears at most once per row and per column. Then if $g \leq L$, the condition C$4$ of Definition~\ref{def-MAPDA} always holds. We only need to prove the case when $g > L$. Consider the case where $f_i \neq f_j$ and $k_i \neq k_j$ for $i \neq j$, with $i,j \in [g]$. From \eqref{eq-k f}, we have
	\begin{align}
		\label{eq-f d}
		\begin{cases}
			2f_1 + d_1=2f_2+d_2, \\
			2f_1 + d_1=2f_3+d_3, \\
			\ \ \ \ \ \ \ \ \ \ \vdots \\
			2f_1 + d_1=2f_g+d_g,
		\end{cases}
		\ \ \text{i.e.,} \ \ 
		\begin{cases}
			d_1-d_2=2(f_2-f_1), \\
			d_1-d_3=2(f_3-f_1), \\
			\ \ \ \ \ \ \ \ \ \ \vdots \\
			d_1-d_g=2(f_g-f_1).
		\end{cases}
	\end{align}
	Without loss of generality, assume that the entries $\mathbf{P}(f_1,k_{i_1}), \mathbf{P}(f_1,k_{i_2}), \ldots, \mathbf{P}(f_1,k_{i_t})$ are all not star, with $t \geq L$, $s \in [t]$, and $i_s \in [g]$. Then by \eqref{eq-k f}, there exist integers $a_1, \ldots, a_t \in [b]$ and $d_{h_s} \in \mathbb{Z}_v$ with $s \in [t]$, $h_s \in [g]$, such that
	\begin{align}
		\label{eq-kit f}
		\begin{cases}
			d_{h_1}=k_{i_1}-f_1 \in \mathcal{D}_{a_1}, \\
			d_{h_2}=k_{i_2}-f_1 \in \mathcal{D}_{a_2}, \\
			\ \ \ \ \ \ \ \ \ \ \vdots \\
			d_{h_t}=k_{i_t}-f_1 \in \mathcal{D}_{a_t}.
		\end{cases}
		\ \ \text{i.e.,} \ \ 
		\begin{cases}
			k_{i_1}=d_{h_1} + f_1, \\
			k_{i_2}=d_{h_2} + f_1, \\
			\ \ \ \ \ \ \ \vdots \\
			k_{i_t}=d_{h_t} + f_1.
		\end{cases}
	\end{align}
	Combining \eqref{eq-k f} and \eqref{eq-kit f} with the relation $k_{i_s} = d_{i_s} + f_{i_s}$ for $s \in [t]$ and $i_s \in [g]$, we obtain
	\begin{align}
		\begin{cases}
			d_{h_1} - d_{i_1}=f_{i_1} - f_1, \\
			d_{h_2} - d_{i_2}=f_{i_2} - f_1, \\
			\ \ \ \ \ \ \ \ \ \ \vdots \\
			d_{h_t} - d_{i_t}=f_{i_t} - f_1.
		\end{cases}
	\end{align}
	From \eqref{eq-f d} and $d_1 - d_{i_s} = 2(f_{i_s} - f_1)$, where $s \in [t]$ and $i_s \in [g]$, it follows that
	\begin{align*}
		\begin{cases}
			d_1 - d_{i_1}=2(d_{h_1}-d_{i_1}), \\
			d_1 - d_{i_2}=2(d_{h_2}-d_{i_2}), \\
			\ \ \ \ \ \ \ \ \ \ \vdots \\
			d_1 - d_{i_t}=2(d_{h_t}-d_{i_t}),
		\end{cases}
		\ \ \text{i.e.,} \ \ 
		\begin{cases}
			d_{h_1} = \frac{d_1 + d_{i_1}}{2}, \\
			d_{h_2} = \frac{d_1 + d_{i_2}}{2}, \\
			\ \ \ \ \ \ \ \vdots \\
			d_{h_t} = \frac{d_1 + d_{i_t}}{2}.
		\end{cases}
	\end{align*}
	This contracts the condition of HSDP that $\sum_{i'=1}^{b}|\mathcal{D}_{i'} \cap \mathcal{B}_{i,j}| < L$. Similarly we can also check other entries. Then the proof is completed.

	\section{Proof of Theorem~\ref{th-main-theorem-MAPDA}}
	\label{proof of Optimal}
	According to Definition~\ref{def-LHSDP}, we first consider the condition that any two difference blocks in $\mathfrak{D}$ are disjoint. Let ${\bf a}=(a_1,a_2,\ldots,a_{n})$, ${\bf a}'=(a'_1,a'_2,\ldots,a'_{n})$, ${\bm \alpha}=(\alpha_1,\alpha_2,\ldots,\alpha_{n+r})$, and  ${\bm\alpha}'=(\alpha'_1,\alpha'_2, \ldots,\alpha'_{n+r})$ be vectors with ${\bf a}, {\bf a} '\in \mathcal{A}$ and ${\bm\alpha},{\bm\alpha}' \in \{1,-1\}^{n+r}$. From \eqref{eq-family} we have two integers of $\mathfrak{D}$, i.e., 
	\begin{align*}
		x=(\alpha_1 a_1x_1+\cdots+\alpha_n a_nx_n)+(\alpha_{n+1}x_{n+1}+\cdots+\alpha_{n+r}x_{n+r}) \in \mathcal{D}_{{\bf a}}
	\end{align*}
	and
	\begin{align*}
		y= (\alpha_1' a_1'x_1+\cdots+\alpha_n' a_n'x_n)+(\alpha_{n+1}'x_{n+1}+\cdots+\alpha_{n+r}'x_{n+r}) \in \mathcal{D}_{{\bf a}'}.
	\end{align*}
	
	Given that $v\geq 2\phi(m_1,m_2,\ldots,m_n)+1=2\sum_{i=1}^{n+r}f(i)+1$, it follows that both $\sum_{j=1}^{n}a_jx_j + \sum_{j=n+1}^{n+r}x_j$ and  $\sum_{j=1}^{n}a_j'x_j + \sum_{j=n+1}^{n+r}x_j$ are less than $\frac{v-1}{2}$. Furthermore, by \eqref{recursive-constr}, for each $i\in[n-1]$ we have $x_{i+1}>2\sum_{j=1}^{i}f(j)$ which implies that
	\begin{align}
		\label{eq-relation_1-n}
		a_{i+1}x_{i+1} > 2\sum_{j=1}^{i}a_jx_j, \ i \in [n-1].
	\end{align} 
	Since $m_{n+1} = m_{n+2} = \dots = m_{n+r} = 1$, for $x_{n+1}$, we have $$x_{n+1} = f(n+1) = f(n) + \frac{f(n)}{m_{n}} = f(n) + 2\sum_{j=1}^{n-1}f(j) + 1 > \sum_{j=1}^{n}f(j) \geq \sum_{j=1}^{n}a_jx_j.$$
	By \eqref{recursive-constr}, for each $i\in[n:n+r-1]$ , we have $x_{i+1} = f(i+1) \geq 2x_{i}$ which implies that
	\begin{align}
		\label{eq-relation_n+1-n+r}
		x_{i+1} > \sum_{j=1}^{n}a_jx_j + \sum_{j=n+1}^{i}x_j, \ i \in [n:n+r-1].
	\end{align}
	When $x=y$, we have $\alpha_{n+r}=\alpha'_{n+r}$. Otherwise, from \eqref{eq-relation_n+1-n+r} with $i=n+r-1$, if $\alpha_{n+r}\neq \alpha'_{n+r}$, without loss of generality, we assume that $\alpha_{n+r}<0$ and $\alpha'_{n+r}>0$. Then we have $\frac{v-1}{2}< x<v$ and $y\leq \frac{v-1}{2}$, which implies $x\neq y$. This contradicts the assumption that $x=y$. So we only need to consider the case $x-\alpha_{n+r}x_{n+r}=y-\alpha'_{n+r}x_{n+r}$. Similarly, from \eqref{eq-relation_n+1-n+r} with $i=n+r-2$, we can obtain $\alpha_{n+r-1}=\alpha'_{n+r-1}$. Using the aforementioned proof method, we can analogously obtain $\alpha_i=\alpha'_i$ for $i \in [n+1:n+r]$. Then we need to consider the case $x-\sum_{i=n+1}^{n+r}\alpha_{i}x_i=y-\sum_{i=n+1}^{n+r}\alpha'_{i}x_i$. If $x-\sum_{i=n+1}^{n+r}\alpha_{i}x_i=y-\sum_{i=n+1}^{n+r}\alpha'_{i}x_i$, we have $\alpha_{n}=\alpha'_{n}$ and $a_{n}=a'_{n}$. Otherwise, from \eqref{eq-relation_1-n} with $i=n-1$, if $a_{n}\neq a'_{n}$ or $\alpha_{n}\neq \alpha'_{n}$, a similar reasoning to the above shows that $x-\sum_{i=n+1}^{n+r}\alpha_{i}x_i \neq y-\sum_{i=n+1}^{n+r}\alpha'_{i}x_i$. By the same reasoning, we can analogously obtain $a_i=a'_i$ and $\alpha_i=\alpha'_i$ for $i \in [n]$. From the above discussion, we conclude that ${\bf a}={\bf a}'$ and ${\bm \alpha}={\bm \alpha}'$. So each integer in $\mathfrak{D}$ appears exactly once.
	
	For any vector ${\bf a}_i=(a_{i,1},a_{i,2},\ldots,a_{i,n})\in \mathcal{A}$, $\mathcal{D}_{{\bf a}_i} = \{d_{i,1}, d_{i,2}, \ldots, d_{i,g}\} \in \mathfrak{D}$ and any vector ${\bm {\alpha}_{i,j}}=({\alpha}_{i_1,j},{\alpha}_{i_2,j},\ldots,{\alpha}_{i_{n+r},j}) \in \{-1, 1\}^{n+r}$ where $i \in [b]$ and $j \in [g]$, we have $$d_{i,j} = ({\alpha}_{i_1,j}a_{i,1}x_1 + {\alpha}_{i_2,j}a_{i,2}x_2 + \ldots +{\alpha}_{i_n,j}a_{i,n}x_n) + ({\alpha}_{i_{n+1},j}x_{n+1} + \ldots + {\alpha}_{i_{n+r},j}x_{n+r}).$$ So the half-sum set of integer $d_{i,j}$, i.e.,
	\begin{align*}
		\mathcal{B}_{i,j} =&\left \{\frac{d_{i,j} + d_{i,1}}{2}, \frac{d_{i,j} + d_{i,2}}{2}, \ldots, \frac{d_{i,j} + d_{i,{j-1}}}{2}, \frac{d_{i,j} + d_{i,{j+1}}}{2}, \ldots, \frac{d_{i,j} + d_{i,{g}}}{2}\right\}\\
		=&\left\{(\frac{{{\alpha}_{i_1,j} + \alpha}_{i_1,1}}{2}a_{i,1}x_1 + \dots + \frac{{\alpha}_{i_n,j} + {\alpha}_{i_n,1}}{2}a_{i,n}x_n) + (\frac{{{\alpha}_{i_{n+1},j} + \alpha}_{i_{n+1},1}}{2}x_{n+1} + \dots + \frac{{\alpha}_{i_{n+r},j} + {\alpha}_{i_{n+r},1}}{2}x_{n+r}), \right.\\
		&(\frac{{{\alpha}_{i_1,j} + \alpha}_{i_1,2}}{2}a_{i,1}x_1 + \dots + \frac{{\alpha}_{i_n,j} + {\alpha}_{i_n,2}}{2}a_{i,n}x_n) + (\frac{{{\alpha}_{i_{n+1},j} + \alpha}_{i_{n+1},2}}{2}x_{n+1} + \dots + \frac{{\alpha}_{i_{n+r},j} + {\alpha}_{i_{n+r},2}}{2}x_{n+r}), \dots, \\
		&\left.(\frac{{{\alpha}_{i_1,j} + \alpha}_{i_1,g}}{2}a_{i,1}x_1 + \dots + \frac{{\alpha}_{i_n,j} + {\alpha}_{i_n,g}}{2}a_{i,n}x_n) + (\frac{{{\alpha}_{i_{n+1},j} + \alpha}_{i_{n+1},g}}{2}x_{n+1} + \dots + \frac{{\alpha}_{i_{n+r},j} + {\alpha}_{i_{n+r},g}}{2}x_{n+r})
		\right\}.
	\end{align*}
	Since $\bm {\alpha}_{i,j} \in \{-1,1\}^{n+r}$ where $i \in [b]$ and $j \in [g]$, it follows that $\frac{{\alpha}_{i_h,j} + {\alpha}_{i_h,l}}{2} \in \{-1, 0, 1\}$ for each $l \in [g]\setminus\{j\}$ and $h \in [n+r]$. Furthermore for each integer $l \in [g]\setminus\{j\}$, there exists at least one integer $h' \in [n+r]$ such that $\frac{{\alpha}_{i_{h',j}} + {\alpha}_{i_{h',l}}}{2} = 0$. Otherwise we have $\bm {\alpha}_{i,l} = \bm {\alpha}_{i,j}$ which implies $d_{i,l}=d_{i,j}$. This contradicts the uniqueness of each integer in $\mathfrak{D}$.
	
	Now let us check the property of $\sum_{i'=1}^{b}|\mathcal{D}_{i'} \cap \mathcal{B}_{i,j}| < L$ where $i \in [b]$ and $j \in [g]$. If the intersection of $\mathcal{B}_{i,j}$ and each block is empty, $\sum_{i'=1}^{b}|\mathcal{D}_{i'} \cap \mathcal{B}_{i,j}| = 0 < L$ always holds. So it is sufficient to consider the case that the intersection of  $\mathcal{B}_{i,j}$ and some block is not empty.
	
	Next, we will prove that when $\mathcal{B}_{i,j}$ intersects with blocks, $\sum_{i'=1}^{b}|\mathcal{D}_{i'} \cap \mathcal{B}_{i,j}| < L$ holds.
	Before proving $\sum_{i'=1}^{b}|\mathcal{D}_{i'} \cap \mathcal{B}_{i,j}| < L$, we first prove that when $\mathcal{B}_{i,j}$ intersects with blocks, the Tail--Zero property holds, i.e., if a half-sum contained in a block, for $s \in [n+1:n+r]$, there exists at least one of the $\frac{{\alpha}_{i_s,l} + {\alpha}_{i_s,j}}{2}$ is zero.
	Without loss of generality, we assume that 
	\begin{align*}
		z = (\frac{{{\alpha}_{i_1,j} + \alpha}_{i_1,l}}{2}a_{i,1}x_1 + \dots + \frac{{\alpha}_{i_n,j} + {\alpha}_{i_n,l}}{2}a_{i,n}x_n) + (\frac{{{\alpha}_{i_{n+1},j} + \alpha}_{i_{n+1},l}}{2}x_{n+1} + \dots + \frac{{\alpha}_{i_{n+r},j} + {\alpha}_{i_{n+r},l}}{2}x_{n+r}), l \in [g]\setminus j.
	\end{align*}
	To simplify the proof, let $\gamma_h = \frac{{\alpha}_{i_h,j} + {\alpha}_{i_h,l}}{2}$ where $h \in [n+r]$, so we have 
	\begin{align*}
		z = (\gamma_{1}a_{i,1}x_1 + \dots + \gamma_{n}a_{i,{n}}x_{n}) + (\gamma_{n+1}x_{n+1} + \dots + \gamma_{n+r}x_{n+r}), l \in [g]\setminus j.
	\end{align*}
	Assume that there exists two vectors ${\bf a}'=(a'_1,a'_2,\ldots,a'_{n})\in \mathcal{A}$ and ${\bm \beta}=(\beta_1,\beta_2,\ldots,\beta_{n+r})\in\{-1,1\}^{n+r}$ such that
	\begin{equation}
		\label{y_eq_z}
		\begin{aligned}
			w &= (\beta_1a'_1x_1+\cdots+\beta_{n}a'_{n}x_{n}) + (\beta_{n+1}x_{n+1}+\cdots+\beta_{n+r}x_{n+r}) \\
			&= (\gamma_{1}a_{i,1}x_1 + \dots + \gamma_{n}a_{i,{n}}x_{n}) + (\gamma_{n+1}x_{n+1} + \dots + \gamma_{n+r}x_{n+r}) = z.
		\end{aligned}
	\end{equation}
	We assume that for any $s \in [n+1:n+r]$, $\gamma_{s} \neq 0$. So there exists at least one integer $s' \in [n]$ such that $\gamma_{s'} = 0$. But similar to the proof of the uniqueness of each integer in $\mathfrak{D}$ introduced above, we can get $\beta_h = \gamma_h$ and $a'_h=a_{i,h}$ for every $h \in [n]$. Then there exists at  least one $\beta_{s'}=0$ for some $s'\in [n]$. This contradicts our definition rule given in \eqref{eq-family}, that is, $\beta_{s'} \in \{-1,1\}$. So for $s \in [n+1:n+r]$, there exists at least one of the $\gamma_{s}$ is zero. 
	
	Finally, we will prove $\sum_{i'=1}^{b}|\mathcal{D}_{i'} \cap \mathcal{B}_{i,j}| < L$ where $i \in [b]$ and $j \in [g]$. From Construction~\ref{cons-LHSDP}, we have $$x_{n+r} = f(n+r) = 2x_{n+r-1} + (2^r-L)x_{n+1} = 2^2x_{n+r-2} + (2^r-L)x_{n+1} = \dots = 2^{r-1}x_{n+1} + (2^r-L)x_{n+1},$$ and $$x_{n+1} = (1 + m_{n})x_{n}.$$ Then \eqref{y_eq_z} can be written as  
	\begin{align*}
		w =& \beta_1a'_1x_1 + \cdots + \beta_{n-1}a'_{n-1}x_{n-1} + \\
		&(\beta_{n}a'_{n} + (1+m_{n})(\beta_{n+1} + 2\beta_{n+2} + \dots + 2^{r-2}\beta_{n+r-1} + (2^{r-1} + 2^r - L)\beta_{n+r}))x_{n} \\
		=&\gamma_1a_{i,1}x_1 + \cdots + \gamma_{n-1}a_{i,n-1}x_{n-1} + \\
		&(\gamma_{n}a_{i,n} + (1+m_{n})(\gamma_{n+1} + 2\gamma_{n+2} + \dots + 2^{r-2}\gamma_{n+r-1} + (2^{r-1} + 2^r - L)\gamma_{n+r}))x_{n} \\
		=&z.
	\end{align*}
	Similar to the proof of the uniqueness of each integer in $\mathfrak{D}$ introduced above, we have 
	\begin{align}
		\label{coefficient_eq_1-n-1}
		\beta_h = \gamma_h, a'_h=a_{i,h}, h \in [n-1],
	\end{align}
	and 
	\begin{equation}
		\begin{aligned}
			\label{eq_n_to_n+r}
			&\beta_{n}a'_{n} + (1+m_{n})(\beta_{n+1} + 2\beta_{n+2} + \dots + 2^{r-2}\beta_{n+r-1} + (2^{r-1} + 2^r - L)\beta_{n+r}) \\
			&= \gamma_{n}a_{i,n} + (1+m_{n})(\gamma_{n+1} + 2\gamma_{n+2} + \dots + 2^{r-2}\gamma_{n+r-1} + (2^{r-1} + 2^r - L)\gamma_{n+r}).
		\end{aligned}
	\end{equation}
	To simplify the proof, let $$u = \gamma_{n+1} + 2\gamma_{n+2} + \dots + 2^{r-2}\gamma_{n+r-1} + (2^{r-1} + 2^r - L)\gamma_{n+r},$$ and $$v = \beta_{n+1} + 2\beta_{n+2} + \dots + 2^{r-2}\beta_{n+r-1} + (2^{r-1} + 2^r - L)\beta_{n+r}.$$
	Then \eqref{eq_n_to_n+r} can be written as
	\begin{align}
		\label{simplify_eq_n_to_n+r}
		\gamma_{n}a_{i,n} - \beta_{n}a'_{n} = (1+m_{n})(v - u).
	\end{align}
	By \eqref{simplify_eq_n_to_n+r}, since $\gamma_h \in \{1,0,-1\}$ where $h \in [n+r]$, we can get
	\begin{align}
		\label{v=u}
		v = u, \gamma_{n} = \beta_{n}, a'_{n} = a_{i,n},
	\end{align}
	or 
	\begin{align}
		\label{v=u+1}
		v = u + 1, \gamma_{n} = 1, \beta_{n} = -1,
		\begin{cases}
			a_{i,n} = 1, a'_{n} = m_{n}; \\
			a_{i,n} = m_{n}, a'_{n} = 1,
		\end{cases} 
	\end{align}
	or
	\begin{align}
		\label{v=u-1}
		v = u - 1, \gamma_{n} = -1, \beta_{n} = 1,
		\begin{cases}
			a_{i,n} = 1, a'_{n} = m_{n}; \\
			a_{i,n} = m_{n}, a'_{n} = 1.
		\end{cases} 
	\end{align}
	From \eqref{v=u}-\eqref{v=u-1}, We have $\gamma_{n} \neq 0$. Otherwise, if $\gamma_{n} = 0$, \eqref{simplify_eq_n_to_n+r} has no solution. Combining this with \eqref{coefficient_eq_1-n-1}, we further obtain $\gamma_h \neq 0$ for all $h \in [n-1]$. This shows that the Tail--Zero Property guarantees that the first $n$ coefficients of the block-intersecting half-sum are all nonzero, while among the remaining $r$ coefficients, at least one must be zero.
	
	Since as fixing $d_{i,j}$ also fixes the range of possible values for $\gamma_h$, i.e., $\gamma_h \in \{-1, 0\}$ or $\{1,0\}$ where $h \in [n+r]$. Without loss of generality, we assume that $\gamma_{n} = 1$, then we have $v = u$ or $v = u+1$. Since $\beta_{h} \in \{-1,1\}^{n+r}$ where $h \in [n+r]$, so there are $2^r$ possible values for $v$, i.e., $v \in \{\pm(2^r-L+1), \pm(2^r-L+3), \pm(2^r-L+5), \dots, \pm(2^r-L+2^r-1)\}$. Let  $L = 2^r-c$ where $c \in [0, 2^{r-1}-1]$, then we have 
	\begin{align}
		\label{range_v}
		v \in \{\pm(c+1), \pm(c+3), \pm(c+5), \dots, \pm(c+2^r-1)\}.
	\end{align}
	Next, let us consider the range of $u$. To simplify the proof, let ${\bm \tau}_u = (\gamma_{n+1}, \gamma_{n+2}, \dots, \gamma_{n+r})$, When $\gamma_{n+r} = 0$, we have
	\begin{equation}
		\begin{aligned}
			\label{range_u_gamma_neq_0}
			&{\bm \tau}_u \in \{\{1,0\}, \{1,0\}, \{1,0\},\dots, \{1,0\}, \{1,0\}, 0\}, u \in \{0,1,2,3,4,\dots, 2^{r-1}-3, 2^{r-1}-2, 2^{r-1}-1\},\\
			&{\bm \tau}_u \in \{\{-1,0\}, \{1,0\}, \{1,0\},\dots, \{1,0\}, \{1,0\}, 0\}, u \in \{-1,0,1,2,3, \dots, 2^{r-1}-4, 2^{r-1}-3, 2^{r-1}-2\},\\
			&{\bm \tau}_u \in \{\{1,0\}, \{-1,0\}, \{1,0\},\dots, \{1,0\}, \{1,0\}, 0\}, u \in \{-2,-1,0,1,2, \dots, 2^{r-1}-5, 2^{r-1}-4, 2^{r-1}-3\},\\
			&{\bm \tau}_u \in \{\{-1,0\}, \{-1,0\}, \{1,0\},\dots, \{1,0\}, \{1,0\}, 0\}, u \in \{-3,-2,-1,0,1, \dots, 2^{r-1}-6, 2^{r-1}-5, 2^{r-1}-4\},\\
			&\ \ \ \ \ \ \ \ \ \  \ \ \ \ \ \ \ \ \ \ \ \ \ \ \ \ \ \ \ \ \ \ \ \ \ \ \  \ \ \ \ \ \ \ \ \ \ \ \ \ \ \ \ \  \ \ \ \ \vdots \\
			&{\bm \tau}_u \in \{\{1,0\}, \{-1,0\}, \{-1,0\},\dots, \{-1,0\}, \{-1,0\}, 0\}, u \in \{-(2^{r-1}-2),-(2^{r-1}-3),\dots,-1,0,1\},\\
			&{\bm \tau}_u \in \{\{-1,0\}, \{-1,0\}, \{-1,0\},\dots, \{-1,0\}, \{-1,0\}, 0\}, u \in \{-(2^{r-1}-1),-(2^{r-1}-2),\dots,-2,-1,0\}.
		\end{aligned}
	\end{equation}
	From \eqref{range_v}, together with $v = u$ or $v = u+1$, a corresponding $v$ can exist only if $u \in \{\pm c, \pm (c+1), \pm (c+2), \dots, \pm (2^{r-1}-1)\}$. Together with \eqref{range_u_gamma_neq_0}, for each value of $\tau_u$, the number of intersections between $u$ and $v$ is at most $2^{r-1}-c$. When $\gamma_{n+r} \neq 0$, we have 
	\begin{align}
		\label{range_u_gamma_eq_0}
		u \in \{\pm(c+1), \pm(c+2), \pm(c+3), \dots, \pm(c+2^r-2)\}, &\ \text{if}\ \gamma_{n+r} \neq 0,
	\end{align}
	where $u$ can only take $2^{r-1}-1$ values, since for $s \in [n+1:n+r]$, there exists at least one of the $\gamma_{s}$ is zero.
	From \eqref{range_v} and \eqref{range_u_gamma_eq_0}, together with $v = u$ or $v = u+1$, we can always find a corresponding $v$, regardless of the value of $u$. In summary, there are at most $2^{r-1}-c+2^{r-1}-1=2^r-c-1$ possible values for u, which makes \eqref{simplify_eq_n_to_n+r} hold. Then for any $d_{i,j} \in \mathcal{D}_{{\bf a}_i}$, with its corresponding $\mathcal{B}_{i,j}$, we have $\sum_{i'=1}^{b}|\mathcal{D}_{{\bf a}_{i'}} \cap \mathcal{B}_{i,j}| \leq =2^r-c-1 < 2^r-c=L$ where $i \in [b]$ and $j \in [g]$. Then the proof is completed.
	
	\section{Proof of Theorem~\ref{th_sub_operation}}
	\label{sec:proof of th_sub_operation}
	
	When $m_1$, $m_2$, $\ldots$, $m_{n}$ are real variables, the problem is a convex $n$-dimensional program over the reals. Its optimal solution can be found via the Lagrange Multiplier Method, by computing the partial derivatives of the function:
	\begin{align*}
		\psi(m_1,\ldots,m_{n}) = &\prod_{i=1}^{n}m_{i}+\lambda\Biggl(\frac{v-1}{2}-\left(2(2^{r+1}-L)(1+m_{n}) -1\right)\left(\sum_{i=1}^{n-2}\left(m_i\prod_{j=i+1}^{n-1}(1+2m_{j})\right) + m_{n-1}\right)-\\
		&\left(2^{r+1}-L\right)\left(1+m_{n}\right)+1\Biggr).
	\end{align*}
	For the variable $m_1$, we have
	\begin{align*}
		\frac{\partial \psi(m_1,\ldots,m_{n})}{\partial m_1} = \prod_{i=2}^{n} m_i - \lambda\left(2(2^{r+1}-L)(1+m_{n}) -1\right) \prod_{i=2}^{n-1}(1 + 2m_i) = 0,
	\end{align*}
	which implies 
	\begin{align}
		\label{eq-differ-m1}
		\prod_{i=2}^{n} m_i = \lambda\left(2(2^{r+1}-L)(1+m_{n}) -1\right) \prod_{i=2}^{n-1}(1 + 2m_i).
	\end{align}
	Similarly, the partial derivative of $\psi(m_1,\ldots,m_{n+r})$ with respect to $m_2$ is
	\begin{align*}
		\frac{\partial \psi(m_1,\ldots,m_{n})}{\partial m_2} = m_1 \prod_{i=3}^{n} m_i - \lambda\left(2(2^{r+1}-L)(1+m_{n}) -1\right) \left( 2m_1 \prod_{i=3}^{n-1}(1 + 2m_i) + \prod_{i=3}^{n-1}(1 + 2m_i) \right) = 0,
	\end{align*}
	which implies 
	\begin{align}
		\label{eq-differ-m2}
		m_1 \prod_{i=3}^{n} m_i = \lambda\left(2(2^{r+1}-L)(1+m_{n}) -1\right) (1+2m_1)\prod_{i=3}^{n-1}(1 + 2m_i).
	\end{align}
	By substituting \eqref{eq-differ-m1} into \eqref{eq-differ-m2} and simplifying, we get $(2m_2+1)m_1=m_2(2m_1+1)$, hence $m_1=m_2$. Similarly, the partial derivative of $\psi(m_1,\ldots,m_{n+r})$ with respect to $m_3$ is
	\begin{align*}
		\frac{\partial \psi(m_1,\ldots,m_{n})}{\partial m_3}=& m_1m_2 \prod_{i=4}^{n} m_i - \lambda\left(2(2^{r+1}-L)(1+m_{n}) -1\right) \\
		&\left( 2m_1(1+2m_2)\prod_{i=4}^{n-1}(1 + 2m_i) +2m_2\prod_{i=4}^{n-1}(1 + 2m_i)+\prod_{i=4}^{n-1}(1 + 2m_i) \right) = 0,
	\end{align*}
	which implies 
	\begin{align}
		\label{eq-differ-m3}
		m_1m_2\prod_{i=4}^{n} m_i = \lambda\left(2(2^{r+1}-L)(1+m_{n}) -1\right)(2m_1(1+2m_2) + 2m_2 + 1) \prod_{i=4}^{n-1}(1 + 2m_i).
	\end{align}
	By substituting \eqref{eq-differ-m1} and $m_1=m_2$ into \eqref{eq-differ-m3} and simplifying, we have $(2m_3+1)m_1=m_3(2m_1+1)$ which implies $m_3=m_1$. 
	
	We now proceed by induction to prove that $m_1=m_2=\cdots=m_{n-1}$. Suppose $m_1=m_2=\cdots=m_{u}$ holds for some $3 \leq u < n-2$. To show that $m_{u+1}=m_1$, we take the partial derivative of $\psi(m_1,\ldots,m_{n})$ with respect to $m_{u+1}$ and obtain
	\begin{equation}
		\begin{aligned}
			\label{eq-differ-u+1}
			\frac{\partial \psi(m_1,\ldots,m_{n})}{\partial m_{u+1}}=&\prod_{i\in [n]\setminus\{u+1\}} m_i-\lambda \left(2(2^{r+1}-L)(1+m_{n}) -1\right)\Biggl(2 \sum_{j=1}^{u-1}\left(m_j\prod_{h\in [j+1:n-1]\setminus\{u+1\}}(1+2m_{h})\right)+\\
			&2m_u\prod_{i=u+2}^{n-1}(1 + 2m_i)+
			\prod_{i=u+2}^{n-1}(1 + 2m_i)\Biggr) = 0.
		\end{aligned}
	\end{equation}
	By submitting the results $m_1=m_2=\cdots=m_u$ into \eqref{eq-differ-u+1}, we have 
	\begin{equation}
		\label{eq-differ-m1_u+1}
		\begin{aligned}
			\prod_{i\in [n]\setminus\{u+1\}} m_i =& \lambda\left(2(2^{r+1}-L)(1+m_{n}) -1\right)\prod_{i=u+2}^{n-1}(1 + 2m_i)\left(2m_1\sum_{j=1}^{u-1}(1+2m_1)^j+1\right)\\
			=&\lambda\left(2(2^{r+1}-L)(1+m_{n}) -1\right)(1+2m_1)^{u}\prod_{i=u+2}^{n-1}(1 + 2m_i).
		\end{aligned}
	\end{equation}
	By substituting \eqref{eq-differ-m1} and $m_1=m_2=\cdots=m_u$ into \eqref{eq-differ-m1_u+1} and simplifying, we have $(2m_{u+1}+1)m_1=m_{u+1}(2m_1+1)$ which implies $m_{u+1}=m_1$. 
	When $u=n-1$, the statement can be verified using the same method. Thus, we conclude that $m_1=m_2=\cdots=m_{n-1}$. Next, considering the variable $m_{n}$, we have  
	\begin{align}
		\label{eq-differ-n}
		\frac{\partial \psi(m_1,\ldots,m_{n})}{\partial m_{n}} = \prod_{i=1}^{n-1}m_{i}-\lambda\Biggl(\left(2^{r+2}-2L\right)\left(\sum_{i=1}^{n-2}\left(m_i\prod_{j=i+1}^{n-1}(1+2m_{j})\right) + m_{n-1}\right)+\left(2^{r+1}-L\right)\Biggr)=0.
	\end{align}
	By submitting the results $m_1=m_2=\cdots=m_{n-1}$ into \eqref{eq-differ-n},  we have 
	\begin{align}
		\label{eq-differ-m1_n}
		\prod_{i=1}^{n-1} m_i = \lambda(2^{r+1}-L)(1+2m_1)^{n-1}.
	\end{align}
	By substituting \eqref{eq-differ-m1} and $m_1=m_2=\cdots=m_{n-1}$ into \eqref{eq-differ-m1_n} and simplifying, we have
	\begin{align*}
		m_1(1+2m_1)^{n-2}\left(2(2^{r+1}-L)(1+m_{n})-1\right) = m_{n}(2^{r+1}-L)(1+2m_1)^{n-1}
	\end{align*}
	which implies 
	\begin{align*}
		m_{n} = \frac{(2^{r+2}-2L-1)m_1}{2^{r+1}-L}.
	\end{align*}
	From the above, the optimal solution to \eqref{eq-optimization} satisfies $m_1=m_2=\cdots=m_{n-1}, m_{n} = \frac{(2^{r+2}-2L-1)m_1}{2^{r+1}-L}$. Combining this with \eqref{eq_sum_f} and \eqref{eq-optimization}, we have
	\begin{align*}
		\left(2(2^{r+1}-L)(1+\frac{(2^{r+2}-2L-1)m_1}{2^{r+1}-L})-1\right)\frac{(1+2m_1)^{n-1}-1}{2} + (2^{r+1}-L)(1+\frac{(2^{r+2}-2L-1)m_1}{2^{r+1}-L})-1 \leq \frac{v-1}{2},
	\end{align*}
	which implies $$m_1 \leq \frac{\sqrt[n]{\frac{v}{2^{r+2}-2L-1}}-1}{2}.$$
	So the total number of blocks of HSDP is $$b \leq \frac{(2^{r+2}-2L-1)(\sqrt[n]{\frac{v}{2^{r+2}-2L-1}}-1)^{n}}{2^{n}(2^{r+1}-L)},$$ 
	and the memory ratio is at least 
	$$1-\frac{2^{n+r}(2^{r+2}-2L-1)(\sqrt[n]{\frac{v}{2^{r+2}-2L-1}}-1)^{n}}{(2^{r+1}-L)v}.$$

	\bibliographystyle{IEEEtran}
	\bibliography{Reference.bib}
\end{document}